\documentclass[aps,pre,twocolumn,nopacs,amssymb]{revtex4}

\usepackage{graphicx}
\usepackage{dcolumn}
\usepackage{bm}
\def\gs{\gtrsim}
\def\ls{\lesssim}
\def\be{\begin{equation}}
\def\en{\end{equation}}    
\def\gs{\gtrsim}
\def\ls{\lesssim}
 
\newcommand{\bi}[1]{\mbox{\boldmath$#1$}}
\newcommand{\av}[1]{\langle{#1}\rangle}
\newcommand{\AV}[1]{{\large\langle}{#1}{\large\rangle}}
\def\p{\partial}
\def\bea{\begin{eqnarray}}
\def\ena{\end{eqnarray}}

\begin{document}
\preprint{APS}
\title{Density functional theory of  gas-liquid phase separation in dilute 
binary mixtures  }

\author{Ryuichi Okamoto$^1$}
\author{Akira Onuki$^2$}
\affiliation{
$^1$Department of Chemistry, Tokyo Metropolitan University, 
Hachioji, Tokyo 192-0397, Japan\\
$^2$Department of Physics, Kyoto University, Kyoto 606-8502, Japan}

\date{\today}

\begin{abstract}  
We examine statics and  dynamics of phase-separated states  
of dilute binary mixtures using density functional theory. 
In our  systems,   the difference in  the   solvation chemical  potential 
$\Delta\mu_s$ 
between  liquid  and gas is considerably 
 larger than the thermal energy $k_BT$  for each   solute particle   
and  the  attractive interaction 
among the solute particles is weaker than that among the solvent  particles. 
 In  these conditions, the saturated vapor pressure 
 increases by an amount equal to the solute density 
in liquid multiplied by  the large factor 
$k_BT \exp(\Delta\mu_s/k_BT)$.  As a result, phase separation 
is induced   at  low solute densities  in liquid 
and the new phase remains in gaseous states,  while    the 
liquid pressure is outside  the coexistence curve of the solvent. 
This explains the  widely 
observed  formation of stable nanobubbles in ambient water 
with a dissolved gas.  We calculate the density and stress 
profiles  across   planar and spherical interfaces, 
where the surface  tension decreases with increasing 
 the interfacial solute adsorption. 
We   realize stable  solute-rich  bubbles   with radius 
about 30 nm, which minimize the free energy functional.   
 We then study   dynamics 
around  such a bubble after a  
decompression of the surrounding liquid, where 
the bubble  undergoes a damped oscillation.
In addition, we  present some exact and approximate 
expressions  for the surface tension 
and the interfacial stress tensor.  
\end{abstract}
\pacs{64.75.Cd, 68.03.-g, 68.08.-p,82.60.Nh}
\maketitle

\section{Introduction}

Much attention has been paid to 
complex interactions among  dissolved particles  
and  solvent molecules  \cite{Likos,Hansen,Hop}.  
In liquid water, hydrophobic particles 
deform the surrounding hydrogen bond structure\cite{Chandler,Garde,Paul}, 
resulting in  a  solvation chemical potential 
$\mu_s$ much larger than the thermal energy $k_BT$ (per particle).
As a result, they tend to  aggregate  in 
 liquid water at ambient conditions (room temperature and 1 atm pressure). 
In simulations of   a hard-sphere particle with radius $a$ in ambient water,  
 $\mu_s$ is of order  $4\pi a^2\sigma$  for $a> 1$ nm, where 
$\sigma$ is the gas-liquid surface tension \cite{Chandler,Garde,Paul}. 
Such large particles are thus strongly hydrophobic 
($\mu_s \sim 180k_BT$  for $a\sim 1$ nm). Another notable example 
is the solvent-mediated  interaction among colloidal particles 
 in near-critical mixture solvents
\cite{Es,Beysens,Dietrich,Oka1,Tanaka,Araki,Yabu,Furu}, 
where one fluid component is preferentially adsorbed 
on the  colloidal surfaces,  largely deforming  the 
surrounding critical fluctuations and often leading to local 
phase separation (bridging).

Recently, we have presented a theory on the formation of small bubbles 
in ambient water containing  a small amount of a dissolved gas  
\cite{nano}.  As a typical example,   O$_2$ is 
  {\it mildly hydrophobic}  with $\mu_s=3.44 k_BT$ in liquid  water  
at $T=300$ K. 
As a unique feature of    ambient liquid water, it  has pressures  
  only slightly higher  than the 
saturated vapor  pressure or is immediately  
 outside the coexistence curve (CX) in the phase diagram. 
Moreover, the van der Waals  interaction among  O$_2$ molecules   
is relatively  weak as compared to the hydrogen bond interaction. 
In fact,  the critical temperature of water is  considerably higher than
that of O$_2$  (647.3 K for water and 154.6 K for O$_2$). 
In this situation, O$_2$  molecules tend to be expelled 
from liquid water in the form of  oxygen-rich bubbles (or films) 
above a very low  threshold  concentration outside CX. 
In many experiments, small bubbles, often  called nanobubbles, 
  have been observed in the bulk and on hydrophobic walls 
in ambient water  \cite{review1,review2}. 
Their radius  is  typically of order $10-100$ nm  
and their life time is   very long. 
In our theory\cite{nano}, they are thermodynamically 
stable, minimizing the free energy including the surface tension. 
 As a similar  phenomenon, 
long-lived heterogeneities have been observed in one-phase states of
aqueous mixtures with addition of a small amount of a salt or a hydrophobic
solute\cite{Ani,Oka-p,Bu}. These phenomena 
emerge as examples  of selective 
solvation effects\cite{Bu}.  

In this paper, we investigate 
 two-phase states  of dilute binary mixtures 
at  $T=300$ K using  the density functional  theory (DFT) 
\cite{Hop,Tara,Sullivan,Evansreview,Lutsko} 
on the basis of   the  Carnahan-Starling 
model for binary mixtures \cite{Car0,Car}. 
As in the case of O$_2$ in water, 
we determine the model parameters in Ref.\cite{Car} such that 
  $\mu_s$ is   $3.44 k_BT$  in liquid and    the  solute-solute 
attractive interaction is weaker than that among the solvent  particles. 
In     these conditions, we  consider  
 gas-liquid coexistence separated by 
 planar and spherical interfaces, where we  calculate  
 the  density and stress profiles   
and the solute-induced deviations of  thermodynamic quantities. 
For our parameter values,  
there also arises a significant interfacial  adsorption 
of the solute, which leads  
to a reduction of the surface tension $\sigma$ in accord with 
 the Gibbs  law \cite{Gibbs}. 
We are interested in  stable   solute-induced   bubbles 
   minimizing   the free energy functional of DFT.  
The  radius of a stable bubble  $R_m$ 
 is larger than the critical radius of nucleation 
$R_c$\cite{Katz,Caupin,Langer,Onukibook}, but remains to be vey small.

Within the scheme of DFT, some attempts have been made 
to describe    dynamics of colloidal particles in solvent 
without \cite{Marconi,Archer-Evans} and with \cite{Lowen,Lutsko,Goddard,Donev} 
the hydrodynamic interaction. 
We  also mention dynamic van der Waals theory 
with  gradient entropy and energy \cite{vanderOnuki}, 
which is a generalization of the original 
van der Waals theory\cite{vander}. Using  this scheme, 
Teshigawara and one of the  present authors (A.O.) 
numerically studied  evaporation and condensation 
in inhomogeneous temperature\cite{Teshi,Teshi1,Teshi2}.    
In their simulations, Tanaka and Araki treated 
 colloidal particles  as   highly viscous droplets  with  
 diffuse interfaces \cite{Tanaka}. Their method has been used 
to  study   hydrodynamics and phase separation 
around colloidal particles \cite{Tanaka,Araki,Furu,Yabu}. 
In this paper, we  present dynamic equations for binary mixtures 
composed of small particles,  where 
DFT and hydrodynamics are incorporated on acoustic and diffusive timescales. 
As an application, we  study    dynamics around a bubble 
 after a decompression of the surrounding liquid.

Bubble dynamics is very complicated, where   hydrodynamics 
and gas-liquid phase transition 
are inseparably  coupled \cite{Plesset,Nepp,Szeri,sonol}.
On short timescales, the pressure balance does not hold 
at the interface and the bubble motions become oscillatory 
accompanied by acoustic disturbances.  
In particular, they have been studied extensively 
  under applied  acoustic field. 
On long timescales, the  bubble growth is governed 
by the thermal diffusion in one-component fluids \cite{Langer,Onukibook} 
and by the slower solute diffusion in  mixtures  \cite{Szeri,nano}. 
In this paper, we aim to study these  characteristic features.

This paper is organized as follows. In Sec.II, we 
will present the background of DFT related to our problem. 
In Sec.III, we will examine two-phase states 
of  dilute binary mixtures 
including a  considerably large solvation chemical potential
in DFT. In Sec.IV, dynamics around a bubble will be 
studied numerically. In Appendix B, we will present some relations 
for the interfacial stress tensor in the exact statistical theory.  
 
\section{Theoretical background} 

\subsection{Free energy  functional }

We treat   a  neutral binary mixture 
in  DFT \cite{Sullivan,Evansreview,Tara,Lutsko}, 
where  the  first species  
is  a solvent and the second  one is  a dilute solute. 
The number densities $ n_1({\bi r})$ and $n_2({\bi r})$ 
are coarse-grained smooth functions  of space. 
There are no  Coulombic and dipolar interactions.   
Hereafter, the temperature $T$ is assumed to be a homogeneous 
constant ($=300$ K)  and the Boltzmann constant is set equal to 1.

In DFT, the   Helmholtz free energy functional 
consists of two parts as  ${\cal F}={\cal F}_h+{\cal F}_a$. The 
first  part  is  of the local form 
 ${\cal F}_h=\int d{\bi r} f(n_1,n_2)$  
with the free energy density, 
\be 
f= T \sum_{i}n_i [\ln(n_i\lambda_i^3)-1]
+ f_h(n_1,n_2)+ \sum_i n_i U_i,
\en 
where $\lambda_i (\propto T^{-1/2})$ is the thermal de Broglie length 
and  the space integral is within the  cell. 
The   $ f_h(n_1,n_2)$ arises from the short-ranged repulsive 
interaction and is taken to assume the binary Carnahan-Starling 
form\cite{Car}. See Appendix A for its details. The  $U_i({\bi r})$ 
is the externally applied potential such as  the wall or gravitational 
potential.  The second part  ${\cal F}_a$ arises from  the attractive 
 interaction and is of the form,     
\be
{\cal F}_a = 
 \frac{1}{2}\int d{\bi r}_1\int d{\bi r}_2
\sum_{i,j}\phi_{ij}(r_{12})n_i({\bi r}_1)
n_j({\bi r}_2).
\en 
Here,   $\phi_{ij}(r_{12})$ is an effective  potential, which is 
 negative and continuously depends  on 
the distance  $r_{12}= |{\bi r}_1-{\bi r}_2|$.  
 Its  space integral should be  finite, so we introduce    
\be 
w_{ij}= -4\pi \int_0^\infty dr r^2 \phi_{ij}(r).
\en

We define  the chemical potentials of the two species as the  
functional derivatives 
$\mu_i({\bi r})= \delta {\cal F}/\delta n_i({\bi r})$. 
For the present  model they are expressed as 
\be
\mu_i= T\ln (n_i\lambda_i^3) + \mu_{hi} + \mu_{ai} + U_i,
\en
where $\mu_{hi} =  {\p  f_h}/{\p n_i} $ is 
the repulsive part and $\mu_{ai}$ is the attractive part 
expressed in the convolution form as 
\be 
\mu_{ai}= \int d{\bi r}_1 \sum_j 
\phi_{ij}(r_1) n_j({\bi r}-{\bi r}_1)  .
\en 
 For homogeneous $n_j$  we have  $\mu_{ai}= -\sum_j  w_{ij} n_j$. 

If variations of $n_i$ are sufficiently slow,  
we can use the gradient  expansion  of ${\cal F}_a$ 
\cite{Evansreview,Cahn,Onukibook} to obtain 
 ${\cal F}_a\cong \int d{\bi r}\sum_{ij} 
[-w_{ij}n_in_j/2 + C_{ij} \nabla n_i\cdot{\nabla n_j}/2]$, where  
\be 
C_{ij}=-\frac{2\pi}{3}  \int_0^\infty dr r^4 \phi_{ij}(r).   
\en 
With this approximation, we  recognize   the relationship between  DFT  
and the original van der Waals theory with the gradient free 
energy\cite{vander}.

In the literature\cite{Evansreview,Tara,Lutsko}, 
 $\phi_{ij}(r)$  has often been set equal to 
 the attractive part of the Lennard-Jones 
potential characterized by the parameters  $d_{ij}$ and $\epsilon_{ij}$. 
 For  $r>2^{1/6}d_{ij}$ it is expressed as 
\be 
\phi_{ij}(r)= 4\epsilon_{ij}[ (d_{ij}/r)^{12}-(d_{ij}/r)^{6}],
\en 
For  $r\le 2^{1/6}d_{ij}$, we define $\phi_{ij}(r)= -\epsilon_{ij}$.  
 In our numerical analysis,  at $T=300$ K,  we set   
\bea 
&& 
\hspace{-8mm}
d_1= d_{11}=3.0~{\rm \AA},~  
d_2= d_{22}=2.48~{\rm \AA},~ 
d_{12}=2.74~{\rm \AA}, \nonumber\\
&&
\hspace{-8mm}
\epsilon_{11}=588.76~ {\rm K},~
\epsilon_{22}=108.53~ {\rm K},~ 
\epsilon_{12}=251.26~ {\rm K}.
\ena 
  Hereafter, we write $d_1= d_{11}$  and  $d_2= d_{22}$  for simplicity.
From Eq.(3) these values lead to  $w_{11}/d_1^3 T =30.80$, 
$w_{22}/d_1^3 T=3.21$, and 
$w_{12}/d_1^3 T=10.02$. Here,    $\epsilon_{22}/\epsilon_{11}=
 0.18$,  so the solute-solute 
attractive interaction is much weaker 
than the solvent-solvent one. 
This is one of the conditions of the solute-induced 
bubble formation\cite{nano}.  
In two-phase  coexistence of the first species, 
the  liquid and gas densities are calculated as 
  $n_\ell= 1.0d_1^{-3}=37$ nm$^{-3}$  and 
$n_g= 2.05 \times 10^{-5}d_1^{-3}= 7.7\times 10^{-4}$ nm$^{-3}$. 
On the other hand, 
in  real ambient  water, 
they are  known to be   $n_\ell\sim  33$ nm$^{-3}$ 
and  $n_g  \sim  10^{-3}$ nm$^{-3}$.

\subsection{Stress tensor in DFT}

We examine the stress tensor 
$\Pi_{\alpha\beta}({\bi r})$  ($\alpha,\beta=x,y, z$) 
in DFT. If  $f_h$ 
is  of the local form, the repulsive   part of 
$\Pi_{\alpha\beta} $ is diagonal as  
$p_h\delta_{\alpha\beta}$ with 
\be 
p_h= n_1\mu_{h1}+n_2\mu_{h2}- f_h.
\en   
However, the attractive pair interaction 
gives rise to off-diagonal  components. 
We   express $\Pi_{\alpha\beta}$  as 
\bea
&& \hspace{-5mm}
\Pi_{\alpha\beta}=(Tn+ p_h) \delta_{\alpha\beta}
- \int d{\bi r}_1 \int d{\bi r}_2 
\sum_{ij} \frac{x_{12\alpha}x_{12\beta} }{2r_{12}}\nonumber\\
&&
\times \phi'_{ij}(r_{12}) 
\delta_s({\bi r}, {\bi r}_1,{\bi r}_2) n_i({\bi r}_1) n_j({\bi r}_2), 
\ena
where  $n=n_1+n_2$,    $\phi'_{ij}= 
d\phi_{ij}/dr$, and $x_{12\alpha}= 
x_{1\alpha}-x_{2\alpha}$. We use    the 
 Irving-Kirkwood 
$\delta$ function\cite{Irving,Kirk,Scho,Onukibook},
\be 
\delta_s({\bi r}, {\bi r}_1,{\bi r}_2)
= \int_0^1 d\lambda\delta ({\bi r}-\lambda, {\bi r}_1 -(1-\lambda){\bi r}_2) ,
\en 
which is nonvanishing only when $\bi r$ 
is on the line segment connecting ${\bi r}_1$ and ${\bi r}_2$. 
 The  integrand is appreciable only if   $|{\bi r}-{\bi r}_1|$ 
and $|{\bi r}-{\bi r}_2|$  are shorter than the potential range. 
The expression (10) is approximate, 
while  the exact  one   in terms of $\delta_s$ \cite{Irving}
will be given in Appendix B.

 Using ${\bi r}_{12}\cdot\nabla 
\delta_s({\bi r}, {\bi r}_1,{\bi r}_2)= \delta({\bi r}-{\bi r}_2) 
-\delta({\bi r}- {\bi r}_1)$ and $\phi_{ij}'(r_{12})x_{12\alpha}/r_{12}= 
\p \phi_{ij}(r_{12})/\p x_{1\alpha}$, we find 
\be 
\sum_\beta \nabla_\beta \Pi_{\alpha\beta}= 
\sum_i n_i \nabla_\alpha (\mu_i-U_i),
\en 
where $\nabla_\beta= \p/\p x_\beta$. 
This relation holds generally   for homogeneous $T$ (even in 
nonequilibrium).  
In equilibrium,  $\mu_1$ and $\mu_2$ are homogeneous, 
which leads to  the mechanical equilibrium condition 
$\sum_\beta \nabla_\beta \Pi_{\alpha\beta}+\sum_i n_i \nabla_\alpha U_i=0$ 
from Eq.(12). For homogeneous $n_i$, 
we have the diagonal form $\Pi_{\alpha\beta}= 
(Tn+ p_h- \sum_{ij}w_{ij} n_i n_j/2)\delta_{\alpha\beta}$ 
as in the van der Waals theory\cite{Onukibook,vander}.

\subsection{Equilibrium in one-dimensional geometry}
We assume that our fluid  is between parallel walls 
separated by $L$. The wall  area   much exceeds  $L^2$. 
The $U_i$ in Eq.(1) is the wall potential. 
 Then, all the physical quantities 
depend only on $z$. We  do not consider  capillary wave fluctuations,  
which are inhomogeneous in the $xy$ plane on large scales. 
They are known to  give rise to 
broadening of the profile \cite{Sti,Evansreview,Grest}.

 \subsubsection{Pressure balance and grand potential }

In equilibrium, the chemical potentials in Eq.(4) are homogeneous 
in the cell  as 
$\mu_1= \mu_1^0$ and $ \mu_2=\mu_2^0$,  
where $\mu_1^0$ and $ \mu_2^0$ are constants.  
 After integration  in the $xy$ plane, the attractive part 
$\mu_{ai}$ in Eq.(5) becomes   
\be 
\mu_{ai}(z) = 2\pi \sum_j \int  dz_1 \Phi_{ij}(z-z_1)  n_j(z_1) ,
\en 
where we define the function, 
\be 
\Phi_{ij}(z)=  \int_{|z|}^\infty dr r \phi_{ij}(r).
\en  
Here, 
 $\int dz \Phi_{ij}(z)= 2 \int_0^\infty dr r^2 \phi_{ij}(r)=-w_{ij}/2\pi$, 
so we have $\mu_{ai}= -\sum_j w_{ij}n_j$ for homogeneous $n_j$. 

From Eq.(12)  the stress balance along the $z$ axis gives   
\be 
\frac{d}{dz}
\Pi_{zz}= -\sum_i n_iU_i', 
\en 
where  
$U_i'=dU_i/dz$ and $\Pi_{zz}$  consists of three parts  as 
\be 
\Pi_{zz}(z)=Tn(z) +p_h (z) +p_a(z).
\en  
From  Eq.(10)  the attractive part $p_a$ is expressed  as   
\be 
p_{a}= 2\pi \int_z^L \hspace{-1.5mm} dz_1\int_0^z
\hspace{-1.5mm}
 dz_2 \sum_{ij} z_{12}\phi_{ij}(z_{12} ) n_i(z_1) n_j(z_2),
\en 
where $z_{12}=z_1-z_2$ and $z_2<z<z_1$ 
in the integrand. Here, use has been made of 
the relation,
\be 
\int d{\bi r}_{1\perp}
\delta_s({\bi r}, {\bi r}_1,{\bi r}_2)=[\theta(z-z_2)-\theta 
(z-z_1)]/z_{12},
\en 
where $d{\bi r}_{1\perp}= dx_1dy_1$ 
and  $\theta(u)$ is the step function (
equal to 1 for $u>0$ and to  0 for $u\le 0$). 
From Eq.(15),  $\Pi_{zz}$  
is homogeneous far from the walls 
and its value is written as $p^0_{\rm b}$. This means that  if  two phases 
are separated by a planar interface far from the walls, 
 the pressures in the bulk two regions 
are commonly given by $p^0_{\rm b}$.

The grand potential $\Omega=\int d{\bi r}\omega$ 
is the space integral of its density $\omega$. In the present 
1D case,  we find 
\bea
\omega &=& f+ \frac{1}{2}\sum_i \mu_{ai}n_i  -\sum_i \mu_{i} n_i \nonumber\\
&=& -\Pi_{zz} + p_a-\frac{1}{2}\sum_i \mu_{ai}n_i .
\ena  
In the second line, use has been made of Eqs.(13),  (16), and (17). 
In the bulk,  both $p_a$ and $\sum_{ij}n_i \mu_{ai}/2$ 
 tend to  $ -\sum_{ij} w_{ij}n_i n_j/2$. 
Hence,   $\omega$  deviates from its bulk value  $-p^0_{\rm b}$ 
only near the interface or  the walls.   


\begin{figure}
\includegraphics[width=1\linewidth]{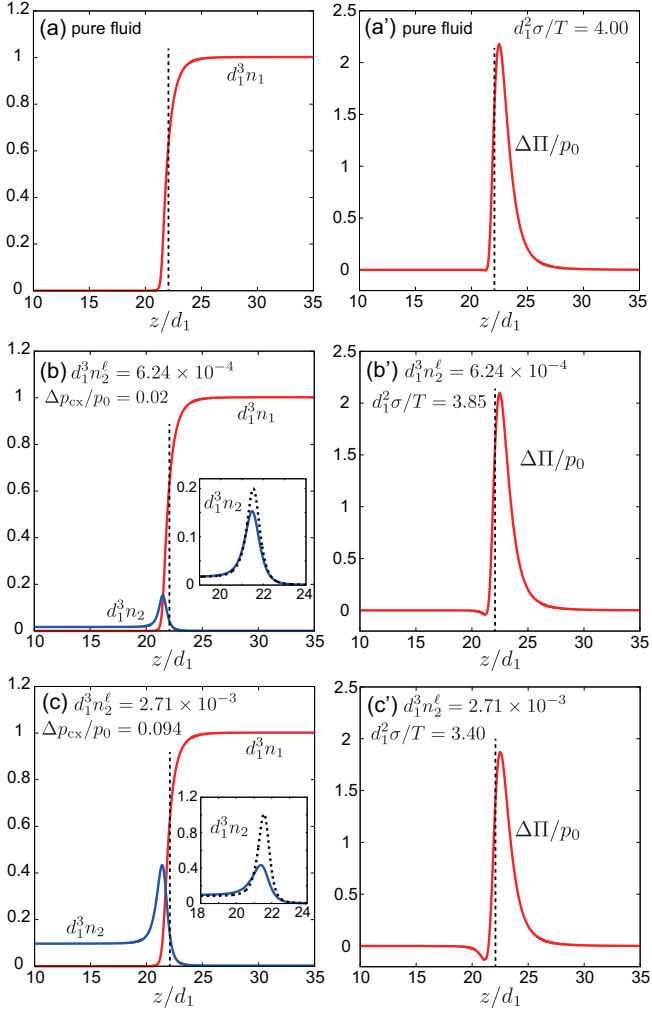}
\caption{ Left: Densities $n_1(z)$ and $n_2(z)$  vs 
$z/d_1$ around 
planar interfaces at $T=300$ K far below 
the  criticality.  Solvent and solute densities  
$(n_1^\ell,  n_2^\ell, n_1^g, n_2^g)$ in bulk liquid and gas are 
  (a) $(1.0, 0, 2.05 \times 10^{-5},0)$  (without solute), 
(b) $(1.0,  6.24 \times 10^{-4}, 2.40 \times 10^{-5}, 2.02 \times 10^{-2})$ 
, and (c) $(1.0, 2.71 \times  10^{-3},4.28 \times 10^{-5},0.097)$ 
 in units of $d_1^{-3}$. In (b) and (c) (inset), 
numerical $n_2(z)$ (bold line) is compared with 
 low density expression in Eq.(37) (dotted line) 
around  the adsorption peak. 
 Right: Stress difference $\Delta\Pi(z)=  \Pi_{zz}(z)- \Pi_{xx}(z)$ 
divided by $p_0= Td_1^{-3}=153$ MPa  vs 
$z/d_1$, calculated from Eqs.(27) and (28) for (a), (b), and (c). 
Its  integral is the surface tension $\sigma$, which  is 
 (a) $ 4.00(=\sigma_0)$, (b) $ 3.85$, 
and (c) $3.40$  in units of  $Td_1^{-2}$.   
The coexisting pressure  $p_{\rm cx}$ is  (a) 
$2.1 \times 10^{-5}(=p_{\rm cx}^0)$, 
(b) $2.0 \times  10^{-2}$, and  (c) $0.094$ in units of 
$Td_1^{-3}$. Vertical broken lines 
indicate the position of the Gibbs dividing surface in Eq.(20). } 
\end{figure}

\subsubsection{ Surface tension}

Let a planar interface parallel to the $xy$ plane 
separate gas and liquid  phases   
far from the walls, where    $U_i= 0$ 
and $\Pi_{zz}=p_{\rm cx}=$const.  around the interface, where $p_{\rm cx}$ 
is the coexisting (saturated pressure) of the mixture.  
The    left panels of Fig.1 display equilibrium 
density profiles $n_1(z)$ and  $n_2(z) $, which 
are  calculated from the  method in Appendix C. 
They  tend to  $n_1^\ell$ and $n_2^\ell$    in  liquid 
($z-z_{\rm int}\gg d_1) 
$and  $n_1^g$ and $n_2^g$    in  gas 
($z_{\rm int}-z\gg d_1)$, respectively.  
 We can  determine the interface position $z_{\rm int}$ by 
\be 
z_{\rm int}= \int_0^L dz [n_1(z)- n_1^g]/[n_1^\ell- n_1^g],
\en 
using the solvent density profile $n_1(z)$\cite{Gibbs}.
 Note that     $n_1^\ell$  is nearly fixed 
 at $1.0d_1^{-3}$ because  the  liquid compressibility is small.

From Eq.(19), 
the surface tension $\sigma$ 
is  the $z$-integral of of $p_a- \sum_i  n_i \mu_{ai}/2$.
Thus,  Eqs.(13) and (17) gives  
\be
\sigma=\frac{1}{2} \int \hspace{-1mm}
dz_1 \int  \hspace{-1mm} dz_2 \sum_{ij}
\Theta_{ij}(z_{12} )  n_i(z_1)n_j(z_2), 
\en
where we define the function,  
\be 
\Theta_{ij} (z)= 2\pi [z^2 \phi_{ij}(|z|)-\Phi_{ij}(z)].
\en  
Here, from  $\int dz \Theta_{ij}(z)=0$, we can replace  
$ n_i(z_1)n_j(z_2)$  by 
$ [n_i(z_1)-n_i(z_2)][ n_j(z_2)-n_j(z_1)]/2$ in Eq.(21); then, 
the integrand is nonvanishing only  near  the interface.
See Appendix B for   the exact 
formula for $\sigma$ \cite{Kirk}. 

Furthermore,  we  introduce  another function $K_{ij}(z)$ by  
\be 
K_{ij}(z)= \frac{\pi}{2} \int^\infty_{|z|}dr r (z^2-r^2)\phi_{ij}(r).  
\en 
This function is related to $\Phi_{ij}$ and $\Theta_{ij}$ as 
\be 
\frac{d}{dz}K_{ij}(z)=\pi z\Phi_{ij}(z),\quad 
\frac{d^2}{dz^2}K_{ij}(z)= -\frac{1}{2}\Theta_{ij}(z).
\en  
In terms of $K_{ij}$ and $n_i'= d n_i/dz$,
 $\sigma$ is expressed as 
\be 
\sigma=\int dz_1\int dz_2\sum_{ij} 
K_{ij}(z_{12})n_i'(z_1)n_j'(z_2).
\en 
This  leads to   the   well-known  form  
$\sigma=\int dz  \sum_{ij} C_{ij} n'_i n_j'$ 
in the gradient theory 
\cite{Onukibook,Cahn,Evansreview}, where  
 $\int dz K_{ij}(z) = C_{ij}$ from 
Eqs.(6) and (23). 
If we assume the Lennard-Jones form in Eq.(7), 
 we have $K_{ij}(z)\sim |z|^{-2}$, 
$\Phi_{ij}(z) \sim |z|^{-4}$, and $\Theta_{ij}(z) \sim |z|^{-4}$ 
 for large $|z|\gg d_1$. 

\begin{figure}
\includegraphics[width=1\linewidth]{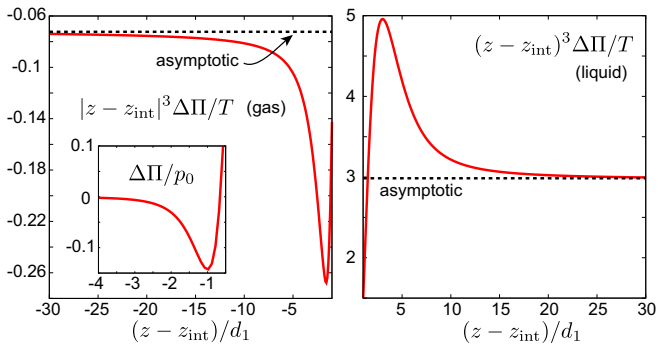}
\caption{ $|z-z_{\rm int}|^{3}\Delta\Pi(z)/T$ 
vs $(z-z_{\rm int})/d_1$ in  gas  (left) 
and liquid  (right) for case (c) in Fig.1 with 
$n_2^\ell= 4.28 \times 10^{-5}d_1^{-3}$. 
It  tends to $-0.0721$(left) and to $ 3.00$ (left) 
in agreement with Eq.(29). In the inset (left) 
$\Delta \Pi/p_0$ is expanded in gas, which has a small 
negative minimum.
}
\end{figure}

The surface tension $\sigma$ of neutral fluids 
can generally be 
expressed by   the Bakker formula\cite{Bakker,Ono,Kirk,Evansreview}, 
\be 
\sigma=\int dz[\Pi_{zz}(z)- \Pi_{xx}(z)]=\int dz \Delta \Pi(z). 
\en 
The stress  difference 
$\Delta\Pi(z)\equiv  \Pi_{zz}(z)- \Pi_{xx}(z)$ 
is nonvanishing only near the interface. 
 In DFT,  we  start with Eq.(10) and use Eq.(18) 
to obtain   
\be 
\Delta \Pi(z)=  \int_z^L \hspace{-0.5mm} dz_1\int_0^z
\hspace{-1.5mm}
 dz_2 \sum_{ij}\frac{\Theta_{ij}(z_{12})}{z_{12}}  n_i(z_1) n_j(z_2), 
\en 
where   $z_1>z>z_2$ in the integrand. 
With respect to $z$,   integration of $\Delta\Pi(z)$ in Eq.(27) 
 gives Eq.(21), while its derivative is written in 
the  single integral form,  
\be 
\hspace{-0.5mm}
\frac{d \Delta\Pi}{dz}=\hspace{-0.5mm} \int_0^\infty \hspace{-1.5mm} 
d\xi \sum_{ij} \frac{\Theta_{ij}(\xi)}{\xi}[ n_i(z+\xi)-n_i(z-\xi)]n_j(z).
\en 
Far from the interface,  
$\Delta \Pi(z)$ decays as  $\propto |z-z_{\rm int}|^{-3}$ 
if the Lennard-Jones potential  in Eq.(7) is assumed. 
In Eq.(28),    we may replace 
$[ n_i(z+\xi)-n_i(z-\xi)]n_j(z)$ by 
$(n_i^\ell-n_i^g)n_j^\ell $  for 
$\xi>z-z_{\rm int}\gg d_1$. Then, for  
$z-z_{\rm int}\gg d_1$, we find  
 $d\Delta\Pi(z)/dz \sim  |z-z_{\rm int}|^{-4}$ and  
\be 
\Delta \Pi (z)\cong \frac{\pi}{2} \sum_{ij} \epsilon_{ij} d_{ij}^6 
(n_i^{\ell}-n_i^g)n_j^\ell  |z-z_{\rm int}|^{-3}. 
\en 
In the gas side with $z_{\rm int}-z\gg d_1$, the corresponding 
tail is obtained if 
$n_j^\ell$ in Eq.(29) is  replaced by $-n_j^g$. Thus, its  amplitude 
 becomes very small in the gas  side. Remarkably, the 
above form of $\Delta\Pi(z)$ 
holds exactly for Lennard-Jones systems (see Appendix B).    
Note  that the density profiles $n_i(z)$ themselves 
 decay as $|z-z_{\rm int}|^{-3}$, as will be shown in Eqs.(43)-(45)  
\cite{Tara,Baker,Hauge}.  In contrast, in   the gradient theory, 
we have  $\Delta\Pi(z)= \sum_{ij} C_{ij} n_i'(z)n_j'(z)$ 
 \cite{Onukibook}, which decays exponentially far from the interface.  

 In the right  panels of Fig.1, 
we  plot  $\Delta\Pi(z)$  calculated from   Eq.(28), 
where its integral ($=\sigma$)  
 decreases with increasing the solute amount 
 due to its  interfacial adsorption 
 (see Fig.5). 
In the liquid side, its  decay  is roughly exponential 
as $\exp[- (z-z_{\rm int})/1.0d_1]$ 
for  $0<z-z_{\rm int}\ls 3d_1$  and is  
algebraic as in Eq.(29) 
 for larger $z-z_{\rm int}$. 
However, in the gas side,  
it decays  rapidly 
and its small  tail is not apparent. 
To detect the tails unambiguously, 
we plot $|z-z_{\rm int}|^{3}\Delta\Pi(z)/T$  in gas and liquid 
in Fig.2.

\subsubsection{ Solid-fluid surface free energy}
 We also derive the expression for  
 the solid-fluid surface free energy $\sigma_{\rm w}$ per unit area, 
which is needed in discussions of the wetting 
and drying transitions \cite{Cahn,Tara,Sullivan}.
 Near the wall at $z=0$, 
 we  find  
$ 
\int_0^\infty dz (\Pi_{zz}-p^0_{\rm b})= 
 \int_0^\infty dz z\sum_i n_i U_i'
$ from   $ d[z\Pi_{zz}]/dz=
 \Pi_{zz} -  z\sum_i n_i U_i' $,   
 where the upper bound  
is pushed to infinity. Then, 
$\sigma_{\rm w}$ is the integral of $p_a - \sum_{i}( \mu_{ai}/2+ zU_i')n_i$ 
so that 
\bea 
\sigma_{\rm w} 
&=& \frac{1}{2}
\int_0^\infty  \hspace{-1mm}
dz_1 \int_0^\infty   \hspace{-1mm} dz_2 
\sum_{ij}   \Theta_{ij}(z_{12})
 n_i(z_1)n_j(z_2)\nonumber\\
&&\hspace{-5mm}
 -\int_0^\infty dz z\sum_i n_i (z) U_i'(z), 
\ena 
where $ \Theta_{ij}(z_{12})
 n_i(z_1)n_j(z_2)$ in the first term can be replaced by 
 $2K_{ij}(z_{12})n_i'(z_1)n_j'(z_2)$ as in Eq.(25). 
 
\section{Dilute binary mixtures}

\subsection{Solvation chemical potential}

Here, we introduce the solvation 
chemical potential for each solute particle 
in  dilute   binary mixtures. The solvation effects are of great importance 
for various solutes including ions 
in aqueous fluids \cite{Bu,Ben,Guillot,Pratt}.

 In the  binary Carnahan-Starling 
model \cite{Car},  the free energy density 
  $f(n_1,n_2)$ in Eq.(1) can be  expanded as 
\be 
 f= f_{\rm w}(n_1)+ Tn_2[\ln (n_2\lambda_2^3) -1 +\nu_h (n_1)],
\en 
up to  order  $n_2$  
\cite{nano,Onuki1}. Here, $f_{\rm w}(n_1)=f(n_1,0)$ is the 
low density limit  of the free energy 
density (excluding the attractive part)  and 
$ \nu_h(n_1)$ arises from the repulsive 
interaction between a solute particle 
and the surrounding solvent. The  solute-solute 
interaction is neglected here. 
As will be shown in Appendix A, we express  $\nu_h$  in terms of 
 $\eta_1=\pi d_1^3 n_1/6$  and  $u_1=\eta_1/(1-\eta_1)$ as   
\bea 
\nu_h&=& (3\alpha+6\alpha^2-\alpha^3 )u_1 + 
(3\alpha^2+4\alpha^3) u_1^2
 \nonumber\\
&&+2\alpha^3u_1^3+ (3\alpha^2-2 \alpha^3-1)\ln \eta_1, 
\ena 
where  $\alpha$ is the solute-to-solvent size ratio,     
\be 
\alpha= d_{22}/d_{11}=d_2/d_1.
\en  
In our numerical  analysis, we set $\alpha=0.827$ in Eq.(8).

Including  the attarctive interaction, the total 
solvation chemical potential $\mu_s$ is written as 
\be 
\mu_s= T\nu_h+ \phi_{12}*n_1.
\en 
where  the second term is of the convolution from. 
The chemical potentials are  approximately given by 
\bea 
&&\hspace{-6mm} 
\mu_1= f_{\rm w}'+\phi_{11}* n_1+  Tn_2 \nu_h' + \phi_{12}* n_2 + U_1,\\
&&\hspace{-6mm} 
\mu_2= T\ln(n_2\lambda_2^3) +\mu_s+U_2,
\ena 
where $f_{\rm w}'= \p f_{\rm w}/\p n_1$ and $\nu_h'= \p \nu_h/\p n_1$.  
We have neglected terms of order  $n_2^2$ in Eq.(35) and those of order  
$n_2$ in Eq.(36). Thus, in equilibrium,   
 $n_2$ is written  as 
\be 
n_2(z) \cong  n_{20} \exp[- \nu_h -\phi_{12}*n_1/T-U_2/T],
\en 
where $n_{20}= \exp(\mu_2/T)\lambda_2^{-3}$. 
In  the insets in Fig.1,  this  
expression and   the numerically 
calculated $n_2(z)$ are compared. The  former noticeably 
overestimates  the adsorption peak in (c), 
where $n_2$ is not small at the peak 
and the repulsive solute-solute interaction 
is appreciable.

\begin{figure}
\includegraphics[width=1\linewidth]{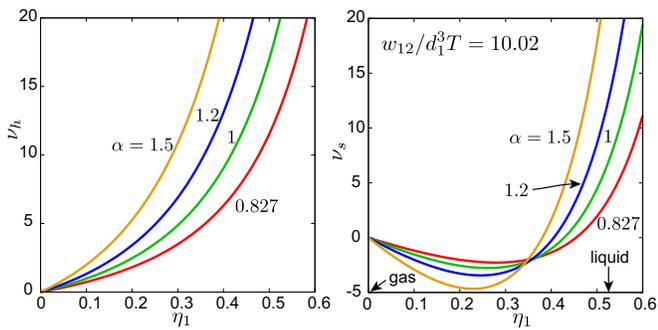}
\caption{ (a) $\nu_h$ in Eq.(32) vs solvent volume fraction $\eta_1=
\pi d_1^3 n_1/6$ for four values of size ratio $\alpha$, which is the 
contribution  from repulsive interaction 
to the solvation  chemical potential   divided by $T$. 
Numerical analysis in this paper is performed for $\alpha=0.827$ 
(lowest curve). (b) Solvation  chemical potential 
 $\nu_s= \nu_h- w_{12} n_1/T $ divided by $T$ 
vs $\eta_1$ including  attractive interaction in bulk with 
$w_{12}/d_1^3T=10.02$.  Marked by arrows indicate  
values of $\eta_1$ in two phase coexistence  
without  solute, where  
 $\nu_s=3.44$  at this liquid point for $\alpha=0.827$. 
 }
\end{figure}

We use the symbol $\nu_s(n_1)$ as the solvation chemical potential 
divided by $T$ in the bulk region:  
\be 
\nu_s(n_1) =\mu_s/T=  \nu_h(n_1)- w_{12}n_1/T.
\en   
whose   derivative  with respect to $n_1$ is written as 
\be 
\nu_s'(n_1) =\nu_h'(n_1)- w_{12}/T.
\en 
The pressure $p$ in the bulk   region 
satisfies  the thermodynamic relation $p= \sum_{i}n_i\mu_i -f-
\sum w_{ij} n_i n_j/2$ as in the van der Waals theory. 
Up to order $n_2$ it  is written  as  
\be 
p=p_{\rm w}(n_1) + T [ 1+ n_1 \nu_s'(n_1)]n_2,
\en
where $p_{\rm w}(n_1)$ is the pressure without solute. 

In Fig.3, we plot $\nu_h$  and $\nu_s$  
vs  $\eta_1=\pi d_1^3 n_1/6$  for four values of $\alpha$. We recognize that 
 $\nu_h$ increases  strongly with increasing 
$\eta_1$ and $\alpha$. In contrast, 
$\nu_s$ in Eq.(38) exhibits a minimum at a   density between the gas and 
liquid densities in the pure fluid limit,  $n_g$ and $n_\ell$. 
Such a   minimum  leads   to solute adsorption in the gas side of 
the interface region (see $n_2(z)$ in Fig.1). 
We set  $w_{12}= 10.02Td_1^3$  
 such that  $\nu_s(n_\ell)-\nu_s(n_g)\cong \nu_s(n_\ell)$ 
is equal to $ 3.44$ (see Eq.(42) below). In our case, we have 
 $n_\ell  \nu_s'(n_\ell) =35.7$, so 
 $\nu_s(n_1)$ is sensitive to  
small variations of $n_1$ and the coefficient of $n_2$ in 
$p$ in Eq.(40) is large in magnitude.

Furthermore, we consider  equilibrium  gas-liquid coexistence,
where the solvent and  solute  densities in the bulk are  $n_1^g$ 
and  $n_2^g$ in gas  and  are  ${n}_1^\ell$ and 
 ${ n}_2^\ell$ in liquid. 
Tnen,  Eq.(37) gives the  solute density ratio,
\be  
{ {n}_2^\ell}/ n_2^g =  \exp ( -\Delta{\nu}_s)=  \exp ( -\Delta{\mu}_s/T).
\en 
Here, $\Delta\mu_s=T\Delta{\nu}_s$ is the difference of 
the solvation  chemical potential  between  
gas and liquid \cite{Ben,Guillot,Bu,Pratt}, which is often called the 
Gibbs transfer free energy (per solute particle in our case). 
 In DFT, Eq.(38) yields  
\be
\Delta\nu_s= [\nu_h(n_1^\ell)-\nu_h(n_1^g)]-(n_1^\ell-n_1^g)w_{12}/T.  
\en  
Note that Eq.(41) is valid  even across curved  interfaces.
If $n_1^\ell\gg n_1^g$, we have $\nu_s(n_1^g)\cong 0$ 
and $\Delta\nu_s\cong \nu_s(n_1^\ell)$.
In the dilute limit 
of solute,  we may replace  $n_1^g$ and $n_1^\ell$ in Eq.(42) by 
their pure fluid  limits, $n_g$ and $n_\ell$, respectively. 
Then, the factor   $\exp(-\Delta\nu_s)$ 
can be related to the Henry constant\cite{Sander,Smith},
from which $\Delta\nu_s$ is $3.44$ for  O$_2$ and 
 is  $4.12$ for N$_2$ in  water  at $T=300$ K\cite{nano}. 
If $\Delta\nu_s$ is much larger, the solute  tends to form  solid  
aggregates in water as hydrophobic hydration  \cite{Chandler,Garde,Paul}. 

\begin{figure}
\includegraphics[width=1\linewidth]{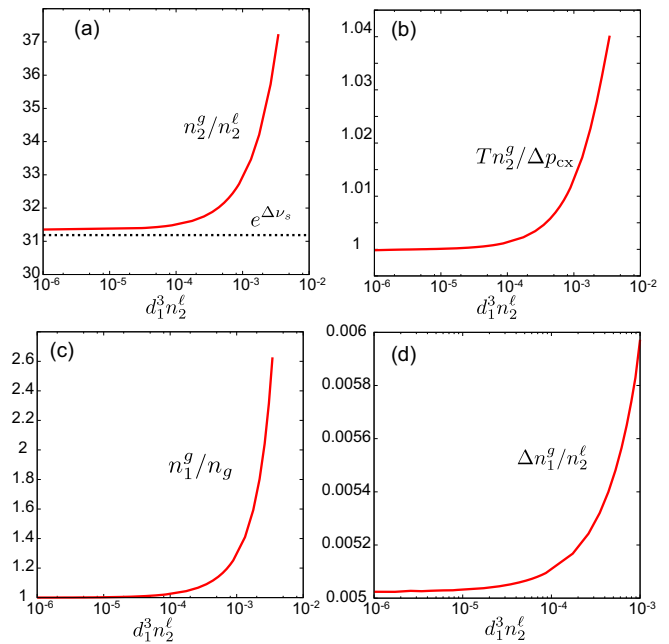}
\caption{ Thermodynamic relations in gas-liquid coexistence 
in dilute binary mixtures with increasing 
solute density $n_2^\ell$ in liquid. 
(a) Ratio $n_2^g/n_2^\ell$ 
of  solute densities in gas and liquid. 
Its dilute limit is 31, but it increases up to 37 
 at $d_1^3 n_2^\ell \sim  0.005$. 
(b) Ratio $Tn_2^g/\Delta p_{\rm cx}$ vs  $d_1^3n_2^\ell$, 
which is very close to 1  supporting Eq.(53). (c) 
Ratio $ n_1^g/n_g $ vs  $d_1^3n_2^\ell$, where 
the pure fluid limit 
$n_g=2.05\times 10^{-5}d_1^{-3}$ 
is very small and $n_i^g$ remains of order $n_g$.  
(d) Ratio $\Delta n_1^g/n_2^\ell$ vs  $d_1^3n_2^\ell$, where 
$\Delta n_1^g=  n_1^g-n_g$. It is about 0.005 in accord with Eq.(51).
 }
\end{figure}

\begin{figure}
\includegraphics[width=1\linewidth]{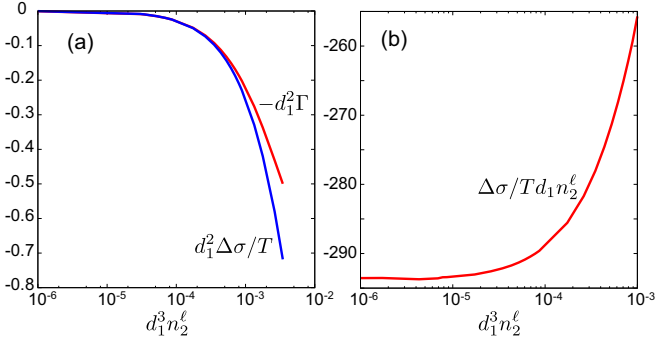}
\caption{ Surface tension change $\Delta\sigma=\sigma-\sigma_0$ 
in dilute  mixtures with increasing 
solute density in liquid $n_2^\ell$. 
(a)  $d_1^2\Delta\sigma/T$ 
 vs  $d_1^3n_2^\ell$, which is compared with normalized 
surface adsorption of solute $d_1^2 \Gamma$ (see  Eq.(57)). 
(b) Ratio $\Delta\sigma/Td_1n_2^\ell$ vs $d_1^3n_2^\ell$, 
which  is about $-290$. }
\end{figure}

 Ishizaki {\it et al.} calculated the solvation chemical potential for 
Lennard-Jones systems via molecular dynamic simulation \cite{Koga}. 
We also note that our solvation chemical potential 
 does not account for the effect of 
orientational degrees of freedom in polar fluids. 
In particular, for water, 
we should investigate  the effect of the 
hydrogen bonding on the solvation chemical potential 
\cite{Borgis}.

\subsection{Algebraic tails in density profiles}

We note that the densities themselves  have 
 algebraic tails for $|z-z_{\rm int}|\gg d_1$ \cite{Tara,Baker,Hauge}, 
as $\Delta\Pi(z)$ in Eq.(29). 
For the Lennard-Jones potential in Eq.(7), Eq.(13) gives 
$\mu_{ai}\cong \sum_j  w_{ij} n_j 
 +T A_i  (z-z_{\rm int})^{-3}$ with 
\be 
A_i= {2\pi} \sum_j \epsilon_{ij} d_{ij}^6 (n_j^\ell-n_j^g)/3T,
\en 
Because  $\mu_i=$const.,      the deviations defined by 
$\delta_\alpha  n_i(z) \equiv  n_i(z)- n_i^\alpha$ ($\alpha=g, \ell$) decay 
for  $|z-z_{\rm int}|\gg d_1$ as 
\bea 
&&\hspace{-5mm}
\delta_\alpha  n_1(z)/n_1^\alpha 
\cong - n_1^\alpha T K_\alpha A_1 /(z-z_{\rm int})^{3},\\
&&\hspace{-5mm}
\delta_\alpha n_2(z)/n_2^\alpha \cong -\nu_s'(n_1^\alpha) \delta_\alpha 
 n_1-  A_2/(z-z_{\rm int})^{3}.
\ena 
In particular, the solute density decays in the gas side 
as $  n_2(z)- n_2^g  \sim  d_{1}^6 n_1^\ell n_2^g |z-z_{\rm int}|^{-3}$. 
We have numerically 
obtained  these tails in excellent agreement with 
Eqs.(42) and (43) (not shown here).

\subsection{Thermodynamics in gas-liquid coexistence}

When gas and liquid phases are separated 
by a planar interface, we consider the 
solute-induced  deviations  in the bulk.  For example, 
the coexisting (saturated vapor) pressure 
  $p_{\rm cx}$ increases with 
increasing $n_2^\ell$ from its pure fluid 
limit  $p_{\rm cx}^0$, where $p_{\rm cx}^0
=2.1 \times 10^{-5}p_0$ with  $p_0=Td_1^{-3}$ in our case. 
To linear order in $n_2^\alpha$, 
the deviation  of $\mu_1$ and  the  shift $\Delta p_{\rm cx}= 
p_{\rm cx}-p_{\rm cx}^0$  are calculated as \cite{Onuki1}  
\bea 
\hspace{-2mm}
\Delta \mu_1&=& (n_\alpha^2 K_\alpha)^{-1} 
\Delta n_1^\alpha + T\nu_s'(n_\alpha) n_2^\alpha, \\
\Delta p_{\rm cx} &=& 
(n_\alpha K_\alpha)^{-1} \Delta n_1^\alpha + 
T[1+n_\alpha\nu_s'(n_\alpha) ] n_2^\alpha \nonumber\\
&=& n_\alpha \Delta \mu_1 +Tn_2^\alpha,  
\ena  
where  $\Delta n_1^\alpha= n_1^\alpha- n_\alpha$ is the solvent 
density deviation and   $K_\alpha$ is the isothermal compressibility 
of pure solvent in phase $\alpha$ defined by 
$f_w''(n_\alpha)- w_{11}=1/n_\alpha^2 K_\alpha$. 
  These relations hold both for gas and liquid $(\alpha=g,\ell)$. Thus,  
\bea 
&& 
\Delta\mu_1= -T[n_2]/[n_1]=
T(n_2^g-n_2^\ell)/(n_\ell-n_g), \\
&&
 \Delta n_1^\alpha= -Tn_\alpha^2 K_\alpha ([n_2]/[n_1]+ 
\nu_s'(n_\alpha) n_2^\alpha ). 
\ena
Here, for any quantity ${\cal A}$,  the difference of the values of ${\cal A}$ 
 in  coexisting  liquid and gas  is written as 
$[{\cal A}]= {\cal A}_\ell-{\cal A}_g$. For example, $[n_i]
= n_i^\ell-n_i^g$. 
 Dividing the second line 
of Eq.(47) by $n_\alpha$, we also  obtain  the solute-induced shift of 
the coexisting   pressure.\cite{Onuki1},  
\be
 \Delta p_{\rm cx}= T\frac{[n_2/n_1]}{[1/n_1]}= T
\frac{n_\ell n_2^g- n_g n_2^\ell}{n_\ell-n_g}.
\en


In our previous paper \cite{nano}
we assumed  the conditions  $n_\ell\gg n_g$ 
and $\exp(\Delta\nu_s)\gg 1$ far from 
the criticality, where  Eqs.(49) and (50) are 
rewritten as 
\bea 
&&\hspace{-5mm}
{\Delta n_1^g} \cong  [1-n_\ell 
  \nu_s'(n_g)]({n_g}/{n_\ell}) n_2^g,\\
&&\hspace{-5mm}
{\Delta n_1^\ell} \cong  {Tn_\ell K_\ell} 
[ 1- e^{-\Delta\nu_s} n_\ell   \nu_s'(n_\ell) ]{n_2^g},\\ 
&&\hspace{-5mm} \Delta p_{\rm cx}\cong Tn_2^g=e^{\Delta\nu_s} Tn_2^\ell,
\ena 
Here, $\Delta n_1^g $ is much  smaller than $n_2^g$ by the factor 
${n_g}/{n_\ell}$  in Eq.(51) 
and the pressure shift $p_{\rm cx}$ 
arises solely from the solute in gas for $n_2^g \gg n_1^g$ in Eq.(53). 
Also   $\Delta n_1^\ell $ can be very small for sufficiently 
small  $K_\ell \ll 1/n_\ell T$.  In our  analysis,  we obtain   
 $Tn_\ell K_\ell=0.025$, 
 $n_\ell \nu_s'(n_g)=-7.07$, 
$n_\ell \nu_s'(n_\ell)= 35.7$, 
 and  $ e^{-\Delta\nu_s} n_\ell  \nu_s'(n_\ell) 
= 1.14$.  Substitution of these values  in Eqs.(51) and (52) yields  
$\Delta n_1^g  \cong 1.6\times 10^{-4}n_2^g$ and 
$\Delta n_1^\ell \cong  -3.6 \times 10^{-3}n_2^g$. Thus,  
 the solvent density is almost  unchanged 
both in  gas and liquid.   


In Fig.4, we  plot the ratios (a)  $n_2^g/n_2^\ell$, 
(b)  $Tn_2^g/\Delta p_{\rm cx}$, (c) 
 $ n_1^g/n_g $, and (d) $\Delta n_1^g/n_2^\ell$ 
as functions of  $n_2^\ell$.    
In (a), the density ratio obeys  Eq.(41)   
 with $\Delta\nu_s=3.44$   for $d_1^3 n_2^\ell\ls 10^{-3}$, but 
it increases up to 37 for larger $n_2^\ell$.  
In  (b),  Eq.(53) holds for all $n_2^\ell$ investigated.  
 In (c),  $n_1^g/n_g$ increases from 1 up to about 2.6  
 staying at very small values.  In (d),     $\Delta n_1^g$ 
increases   with increasing $n_2^\ell$ linearly 
as $\Delta n_1^g \cong 0.005n_2^\ell\cong 1.6\times 10^{-3} n_2^g$ 
in accord   with Eq.(51).

We note that Eqs.(46)-(53) are general thermodynamic relations. 
Let us consider   the Gibbs-Duhem relation 
of binary mixtures at fixed $T$ in  two-phase coexistence,
 \be
dp= n_1^\alpha  d\mu_1+ n_2^\alpha d\mu_2,
\en 
which holds both for $\alpha=g$ and $\ell$. 
Since $n_2 d\mu_2 \cong T dn_2$ for small $n_2$, 
integration of Eq.(54) with respect to  $n_2$ 
yields Eqs.(47)-(50).  
 
With addition of a solute, 
 a homogeneous liquid (without bubbles) 
becomes  metastable against  bubble formation 
if its pressure   $\bar p$  is made slightly lower than 
$p_{\rm cx}= p_{\rm cx}^0+ \Delta p_{\rm cx}$.  
This condition can be realized  even outside the solvent 
coexistence curve ${\bar p}>p_{cx}^0$ if	
$n_2^\ell$ exceeds  a threshold solute density 
$n_2^c$  \cite{nano} given  by 
\be  
 n_2^c =e^{-\Delta\nu_s} ({\bar p}-p_{\rm cx}^0)/T. 
\en  
For ${\bar p}<p_{cx}^0$ (inside CX), 
bubbles can appear even without solute,  
so we may set $n_2^c=0$. 
In ambient water with ${\bar p}-p_{\rm cx}^0\sim 1$ atm, 
$n_2^c$  is very  small even for 
{\it mildly hydrophobic } gases such as O$_2$ and N$_2$\cite{nano}.


We also examine the deviation 
of the surface tension $\Delta\sigma=\sigma-\sigma_0$ 
due to  dilute  solute, 
where $\sigma_0$ is the surface tension without solute.
We consider the deviation of  $\omega$ in Eq.(19), where  
 the coefficient in front of the solute density 
deviation $\Delta n_1(z)$ vanishes   from the homogeneity of $\mu_1$
\cite{Onuki1}. To first order in $n_2$, we find  
\be 
\Delta\sigma= \int dz[ \Delta p_{\rm cx}- n_1(z)\Delta\mu_1- Tn_2(z)]. 
\en 
From Eqs.(48) and (50), we obtain  the Gibbs adsorption formula 
 $\Delta\sigma= -T \Gamma$ \cite{Gibbs}. Here,  $\Gamma$ is 
 the solute  adsorption, 
\be 
\Gamma= \int dz\bigg[ n_2(z)- n_2^g- 
\frac{[n_2]}{[n_1]}(n_1(z)-n_1^g)\bigg]. 
\en 

In Fig.1, we notice  that the adsorption 
occurs in the gas side  in (b) and (c) of Fig.1 
(left  of the Gibbs dividing surface), where  
 $\Gamma= 0.132 d_1^{-2}$ in (b) and 
$\Gamma= 0.432 d_1^{-2}$ in (c). 
Furthermore, in Fig.5, we plot  (a) $\Delta\sigma$ 
and $\Gamma$ and (b) $\Delta\sigma/n_2^\ell$ 
as functions of  $n_2^\ell$. In addition to the 
Gibbs adsorption law, we can see  
the linear behavior  $-\Delta\sigma/d_1T  
\cong 290  n_2^\ell\cong 9.3 n_2^g$ 
for small $n_2^\ell$.

In our previous thermodynamic theory\cite{nano}, we assumed a constant $\sigma$ independent of the solute density. However,   a decrease in 
$\sigma$ increases the stability of small bubbles with a smaller Laplace 
pressure. The Gibbs law  has been used for interfacial adsorption of 
neutral  surfactants. Its 
 generalization including the electrostatic 
interaction was given in our previous papers\cite{Bu,Onukisurf}.

\subsection{Stable solute-induced bubbles}
\begin{figure}
\includegraphics[width=1\linewidth]{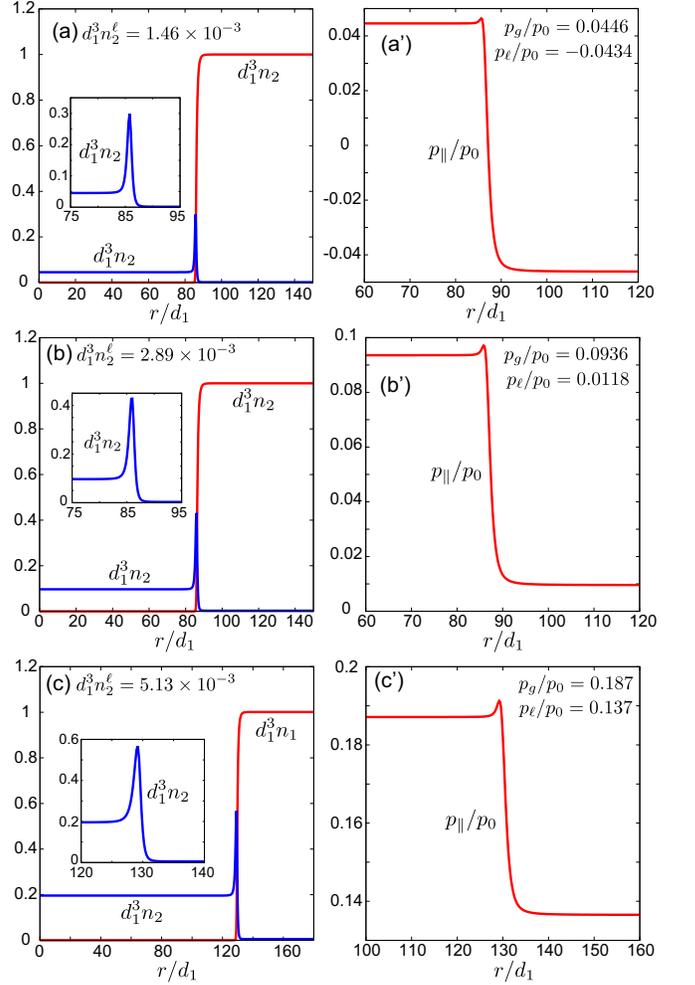}
\caption{ Densities $n_1(r)$ and $n_2(r)$ (left) 
 and pressure $p_\parallel(r)$ along $\hat{\bi r}$ (right) 
 around stable spherical bubbles, where 
 $d_1^3 n_2^\ell$ is (a)  $1.46\times 10^{-3}$, 
(b) $2.89 \times 10^{-3}$, (c) $5.13\times 10^{-3}$, 
with $d_1^3 n_1^\ell\cong 1$. 
Pressures $(p_g, p_\ell)$ in bulk 
gas and liquid   are  (a) $(0.0446,-0.0434) $, 
(b) $(0.0936,0.0118) $,  and (c) $(0.187,0.137) $ in units 
of $p_0=T/d_1^3=153$ MPa, where the difference  
$p_g-p_\ell $ is close to $ 2\sigma/R$.  
Here, $p_\ell$ is negative ($-6$ Mpa) 
 in (a), while it is much above $p_{\rm cx}^0$ for (b) and (c). 
Cell length $L$ is $400d_1$ in (a) and (b) 
and is   $800d_1$ in (c).}
\end{figure}
\begin{figure}
\includegraphics[width=1\linewidth]{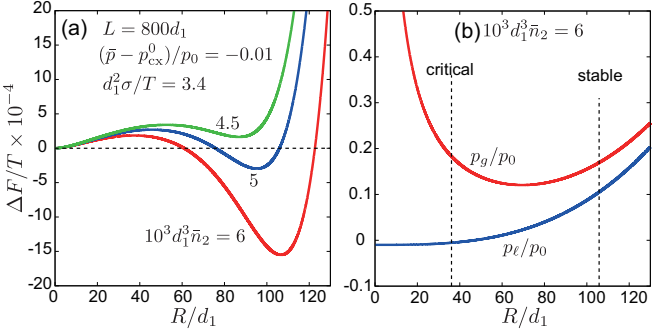}
\caption{ (a) Bubble free energy $\Delta F $ in Eq.(65) vs $R$ 
at fixed volume with $L=800d_1$, where  initial 
pressure is ${\bar p}= p_{cx}^0
-0.01 Td_1^{-3}$. Initial solute density 
is ${\bar n}_2 =4.5, 5$, and $6$ 
in units of $ 10^{-3}d_1^{-3}$ for three curves. 
A stable bubble is realized for the lower two curves at the minimum 
radius $R_m$. (b) Pressures inside and outside bubble $p_g$ and $p_\ell$  
in Eq.(65) for the lowest curve in (a). 
 Vertical broken lines indicate  
critical and stable bubble radii,  
$R_c=37.2 d_1$ and $ R_m=106.6d_1$.  }
\end{figure}

In our previous paper\cite{nano}, we showed  
that  small, stable bubbles   minimize  
an appropriately defined  bubble free energy 
for ${\bar p}>p_{\rm cx}^0$.  They are 
induced by a small amount of 
moderately hydrophobic gas in  water.   
In contrast, in pure fluids,  equilibrium bubbles 
with macroscopic sizes  appear at fixed cell volume 
 inside CX \cite{Onukibook,Binder}. 
Here, we place  a spherical  bubble at the center of  
 a spherical cell,  fixing  the  total 
solvent and solute numbers $N_1$ and $N_2$. 
See  Fig.6 for typical density profiles around stable bubbles.  



\subsubsection{Stress tensor around a bubble and Laplace law} 

In our spherically symmetric geometry, 
we  take the reference frame with the origin  at the bubble center. 
The average stress  tensor is generally  of the form\cite{Ono,Scho}, 
\be 
{\Pi_{\alpha\beta}}=  p_\parallel (r)  \delta_{\alpha\beta} - 
\Delta\Pi (r) (\delta_{\alpha\beta}-{\hat x}_\alpha \hat{x}_\beta),
\en 
where ${\hat{\bi r}}=(\hat{x}, \hat{y},\hat{z})= 
 r^{-1}{\bi r}$.  The parallel component 
$p_\parallel (r)= \sum_{\alpha\beta}{\Pi_{\alpha\beta}}\hat{x}_\alpha 
\hat{x}_\beta$ and the stress difference 
$\Delta\Pi(r)$ depend only on $r$. With Eq.(58), the mechanical 
 equilibrium condition $\sum_\beta \nabla_\beta {\Pi_{\alpha\beta}}=0$ 
is rewritten as  
\be 
\frac{d}{dr}p_\parallel(r) = -\frac{2}{r}  \Delta \Pi(r).
\en

We assume that the bubble radius $R$ much exceeds the molecular lengths, 
where it is determined  by 
\be 
R=\int_0^L  dr  [n_1(r)-n_1^\ell]/[n_1^g-n_1^\ell]. 
\en 
We first calculate $\Delta \Pi (r) $ for   $R\gg  
d_1$.  Since $2\Delta\Pi= 3\sum_{\alpha\beta} 
\Pi_{\alpha\beta}{\hat x}_\alpha \hat{x}_\beta- 
\sum_\alpha \Pi_{\alpha\alpha}$ from Eq.(58),  Eq.(10) yields   
\bea
&& \hspace{-5mm}
\Delta \Pi(r)=\frac{1}{4} 
 \int d{\bi r}_1 \int d{\bi r}_2 
\sum_{ij}[r_{12}- {3}(\hat{\bi r}\cdot {{\bi r}}_{12})^2/r_{12}] 
 \nonumber\\
&&
\hspace{4mm} 
\times  \phi'_{ij}(r_{12}) 
\delta_s({\bi r}, {\bi r}_1,{\bi r}_2) n_i({r}_1) n_j({r}_2), 
\ena 
where  the integrand is nonvanishing only for $r_1<r<r_2$ 
or  $r_2<r<r_1$ due to the $\delta_s$ function. Furthermore, 
if we set $u= r_{12}=|{\bi r}_1-{\bi r}_2|$, the integrand  is also 
nonvanishing  only for $|r_1-r_2|<u<r_1+r_2$. 
Under these conditions of $r_1$ and $r_2$, the angle integration in Eq.(61) 
can be performed with the aid of the relation, 
\be 
\int d\Omega_1\hspace{-1mm}
\int d\Omega_2 
\delta (r_{12}-u) \delta_s({\bi r}, {\bi r}_1,{\bi r}_2) 
= \frac{4\pi u}{rr_1r_2{W}},
\en 
where $d\Omega_i$  ($i=1,2)$  are  the solid angle 
elements in  $d{\bi r}_i=dr_i r_i^2d\Omega_i$     and  
 $W^2 = u^4- 2u^2(r_1^2+r_2^2-2r^2)+(r_1^2-r_2^2)^2$. 
For $r\gg d_1$, we may set  $\hat{\bi r}\cdot {{\bi r}}_{12}
\cong r_1-r_2 $ in Eq.(61) and $ W\cong 2r |r_1-r_2|$  in Eq.(62) to find   
\be
\hspace{-0.8mm} 
\Delta \Pi\cong  \hspace{-1mm}
 \int_z^L \hspace{-1.5mm} dr_1\int_0^z \hspace{-1.5mm}
 dr_2 \sum_{ij} \frac{\Theta_{ij}(r_{1}-r_2)}{r_1-r_2} 
 n_i(r_1) n_j(r_2), 
\en 
which is of the same form as $\Delta\Pi(z)$ in Eq.(27). 
Thus,  $\Delta\Pi(r)$ behaves in the  same manner as $\Delta\Pi(z)$ 
and its $r$-integral is equal 
to  the surface tension $\sigma$ with a correction of order $R^{-1}$. 
The  Laplace law  then follows if 
 Eq.(59) is integrated  across the interface at $r\cong R\gg d_1$.

In Fig.6, we display  $n_1(r)$, $n_2(r)$,  and  $p_\parallel(r)$  
 around stable spherical bubbles, where we calculate the densities 
with the method in Appendix C and   $p_\parallel(r)$  
 from Eqs.(59) and (63). Here, 
$d_1^3n_2^\ell $ is 
(a)  $1.46\times 10^{-3}$, 
(b) $2.89 \times 10^{-3}$, and (c) $5.13\times 10^{-3}$, 
while  $d_1^3 n_1^\ell\cong 1$. 
The cell radius $L$ is 400 in (a) and (b) and is 800 in (c). 
Remarkably,  the liquid pressure $p_\ell$ is negative at  
$-0.0434 p_0= -6$ MPa in (a), 
while $p_\ell/p_0$ is $0.0118$ in  (b) and 0.196 in (c).
A small peak of $p_\parallel$ at the interface in (b) and (c) is due to 
the solute adsorption.  
Then, $R/d_1$ is (a) 86.3,  (b) 86.5, and  (c) 129.8, 
while $d_1^2 \sigma/T$ is (a) 3.67,  (b) 3.40, and (c) 3.06  
from  the integral of $\Delta\Pi$ in Eq.(63). 
These values yield the normalized 
Laplace pressure $2\sigma/Rp_0$ 
as (a) $0.0851$,  (b) $0.0786$,  and (c) $0.0471$, 
in good agreement with the normalized pressure difference 
$(p_g-p_\ell)/p_0$ 
given by (a) $0.088$,  (b) $0.082$, (c) $0.050$, 
where $p_0=Td_1^{-3}=153$ MPa.

Using DFT,  Talanquer {\it et al.}\cite{Ox} 
calculated the density profiles of unstable  critical bubbles  
at $R=R_c$,   where the solute is accumulated in 
the bubble interior  and its density exhibits a mild maximum at the 
interface.  In their molecular  dynamics simulation, 
 Yamamoto and Ohnishi \cite{Yama}   realized 
 stable  helium-rich nanobubbles  in  water. 
They fixed the cell volume to find  slightly negative   
 pressures  in the liquid region as in our Fig.6(a).

\subsubsection{Equilibrium conditions and critical radius}

We start with  a reference (metastable) liquid state without 
 bubbles, where  the densities 
are   ${\bar n}_i= N_i/V$ and the pressure is $\bar p$. 
With appearance of a single 
bubble at  fixed cell volume, the  densities  in liquid are  changed as  
\bea 
n_1^\ell&=& (1+\phi) {\bar n}_1 , \\ 
n_2^\ell&=& {\bar n}_2- \phi n_2^g, 
\ena 
where  $\phi=4\pi R^3/3V$ is  the bubble volume fraction. Here, 
we  assume $\phi <{\bar n}_2/n_2^g \ll 1$. We also  neglect small 
density heterogeneities around the bubble when it is growing 
or shrinking.  
Using   the liquid compressibility $K_\ell$, we write 
the pressures in liquid and gas as 
\bea 
p_\ell&=& {\bar p} + \phi/K_\ell,\\
p_g&=&  {\bar p} + \phi/K_\ell + 2\sigma/R . 
\ena 
In $p_\ell$ we neglect the second term ($\propto n_2$) 
in Eq.(40),  which is allowable 
for $n_2^g/n_\ell \ll [Tn_\ell 
K_\ell (1+n_\ell\nu_s'(n_\ell)]^{-1}\cong 1$. 
 For small  $K_\ell$, the compression pressure 
$\phi/K_\ell$  can be significant even for small $\phi$. 
For  $p_g >p_{\rm cx}^0$, the  solvent density in gas $n_1^g$ 
is almost unchanged from $n_g= p_{\rm cx}^0/T$.
The gas pressure is also given by $p_g= T(n_1^g+n_2^g)$, so    
\be 
n_2^g = ({\bar p}-p_{\rm cx}^0 +2\sigma/R+ \phi/K_\ell)/T.
\en 
If ${\bar p}<p_{\rm cx}^0$, we require  
$p_{\rm cx}^0-{\bar p}<(8/3)(2\pi \sigma^3/ K_\ell V)^{1/4}= 31.4 
T d_1^{-9/4} V^{-1/4}$ to ensure $n_2^g>0$.

 If we further assume 
the chemical balance in Eq.(41), 
$n_2^g$ is also expressed as
\be  
n_2^g= {\bar n}_2/[e^{-\Delta\nu_s}+\phi].
\en 
 For   given ${\bar n}_2$ 
and $\bar p$,  Eqs.(68) and (69) constitute a closed set 
of equations of  $R$.  
Previously \cite{nano}, we solved them  without 
 $\phi/K_\ell$ at fixed ${\bar p}$ in liquid.  
Also in the present fixed-volume 
case, we obtain  two solutions at 
$R=R_c$ and $R_m$ for sufficiently large ${\bar n}_2$, 
as  in  Fig.7(a). The  larger one  $R_m$ 
is   the radius of  a stable or metastable 
bubble, while  the smaller one  $R_c$ is 
the critical radius of an unstable bubble.  
In the limit $\phi\to 0$, $R_c$ is written as 
\bea 
R_c&=&2\sigma/[ T{\bar n}_2e^{\Delta\nu_s}-{\bar p}+p_{\rm cx}^0]\nonumber\\
&=& 2\sigma e^{-\Delta\nu_s}/T({\bar n}_2-n_2^c),
\ena 
where $n_2^c$ is the threshold solute density in Eq.(55).
Thus, solute-induced metastability is realized  for ${\bar n}_2>n_2^c$.
For pure fluids, $R_c= 2\sigma/(p_{\rm cx}^0-{\bar p})$ 
inside CX \cite{Katz}. Azouzi {\it et al.} \cite{Caupin} 
performed a nucleation experiment at 
negative pressures about  $ -100$ MPa. 

\subsubsection{Bubble free energy 
at fixed  $N_1$-$N_2$-$V$-$T$} 

We   set up a  bubble free energy 
$ F_b$, removing  the solvent degrees of freedom. 
At fixed $V$, it is the change of the 
Helmholtz free energy with appearance a single bubble. 
Using  Eqs.(64) and (66), 
we  express it  as \cite{nano} 
\be 
 F_b =V( \phi \omega_g+  \omega_\ell )+S\sigma ,
\en 
where $S=4\pi R^2$,   $\phi=4\pi R^3/3V \ll 1$,  and 
\bea 
 \omega_g &=&  T n_{2}^g  [\ln({n_{2}^g/{\bar n}_{2}})   
   - 1-{\Delta {\nu}_s}] + {\bar p}-p^0_{\rm cx},\\
{\omega}_\ell &=&   T[{n}_{2}^\ell  \ln
({n}_{2}^\ell/{{\bar n}_{2}})+\phi n_2^g ] + \phi^2/2K_\ell . 
\ena  
Here, we assume  Eq.(65); then,  $ F_b$ depends on $n_2^g$ and $\phi$.  
Against infinitesimal changes $n_2^g\to n_2^g+\delta n_2^g$ and 
$\phi\to \phi+\delta\phi$,   the incremental change in $\Delta F$  is 
written as  
\bea 
{\delta F_b} &=&VT [ \ln ({n_{2}^g}/{{n}_{2}^\ell}) 
-\Delta\nu_{\rm s}] \delta(n_{2}^g\phi) \nonumber\\
&&+  V[{p_g-p_{\rm cx}^0} - T n_{2}^g]\delta\phi. 
\ena   
in terms of  $p_g$ in  Eq.(67). In  equilibrium, 
$F_b$ is  minimized with respect 
to $n_{2}^g$ and $\phi$, leading to    
Eqs.(44)  and  (68).

Let us assume the mechanical equilibrium condition (68) 
and not the chemical one  (44).  Note that the former 
is  instantaneously realized  in the slow nucleation process.  
Then, $F_b$  is a  function of $R$ and  
its derivative   is given by  
\be 
\frac{1}{S}\frac{d  F_b}{dR}=\bigg[ \ln \frac{n_{2}^g}{{n}_{2}^\ell} 
-\Delta\nu_{\rm s}\bigg]\bigg 
[{\bar p}-p_{\rm cx}^0 +  
\frac{4\sigma}{3R}+ \frac{4\phi}{3K_\ell}\bigg],  
\en 
which vanishes under Eq.(41). 
  For sufficiently large 
${\bar n}_2$, $F_b(R)$ exhibits  a local maximum at $R=R_c$ 
and a local minimum at $R=R_m (>R_c)$. 
 Here, $R_c$ is the critical radius expressed as in Eq.(70) 
in the limit $\phi\to 0$, 
while $R_m$ is a stable (metastable) 
bubble radius for negative (positive) $ F_b(R_m)$.

In Fig.7(a), we plot  $ F_b(R) $ in Eq.(71) 
vs  $R$ at  $L=800d_1$, where 
 ${\bar n}_1\cong 1$ and 
 ${\bar n}_2 =4.5$, $5$, and $6$ 
in units of $ 10^{-3}d_1^{-3}$ for the three curves. 
 The initial pressure is  ${\bar p}= p_{cx}^0-0.01 p_0$.  
In Fig.7(b), we plot  the pressures 
$p_\ell$ and $p_g$ in Eqs.(66) and (67) vs $R$ for 
the curve of the largest ${\bar n}_2$ in (a).  
The density profiles for the stable bubble 
in this case are close to those  in Fig.6(c).

\section{Dynamics of a solute-induced bubble}    

Finally, we examine  bubble dynamics  
combining DFT and   hydrodynamics. 
In this paper, we assume homogeneous  $T$ 
in the limit of fast heat conduction. 
This approximation is allowable for 
 slow solute diffusion\cite{nano}. 
However, $T$ becomes inhomogeneous 
around  growing or shrinking 
bubbles firstly due to adiabatic heating or cooling of bubbles and secondly 
due to latent heat  in evaporation and condensation   
 \cite{Szeri,Plesset,Nepp,Teshi,Teshi1,Teshi2}. 
We should examine these effects  in  future simulations.

\subsection{Hydrodynamic equations}

We consider the mass densities $\rho_i = m_i n_i$, 
 the velocity field $\bi v$, and 
the momentum density ${\bi J}= \rho{\bi v}$, 
where $m_1$ and $m_2$ are the molecular masses 
and $\rho=\rho_1+\rho_2$ is the total mass density. 
 The mass conservation yields  \cite{Landau} 
\bea 
&&\frac{\p \rho_1}{\p t}= -\nabla (\rho_1{\bi v}-{\bi I}),\\
&&\frac{\p \rho_2}{\p t}= -\nabla (\rho_2{\bi v}+{\bi I}),
\ena
where ${\bi I}$ is the diffusion flux of the form,
\be 
{\bi I}=-  \Lambda \nabla (\mu_2/m_2-\mu_1/m_1).
\en 
If the solute  is  dilute or $n_2$ is small, 
the kinetic coefficient $\Lambda$ is 
related to the solute  diffusion constant $D$ by 
\be 
\Lambda= D m_2^2 n_2/T.
\en 
Then, as $n_2 \to 0$, 
we have $\nabla \rho_2 \cong T n_2^{-1}\nabla n_2$ 
and  ${\bi I}\cong -D\nabla\rho_2$. 
Next, the  momentum equation is written as 
\bea 
&&\hspace{-1cm}
\frac{\p{\bi J}}{\p t}+ \nabla\cdot[\rho{\bi v}{\bi v}]= 
- \nabla \cdot(\tensor{\Pi}-{\tensor\sigma}_{\rm vis})
-\sum_i n_i\nabla U_i 
\nonumber\\
&&\hspace{1.4cm}=
- \sum_i n_i\nabla \mu_i +\nabla\cdot{\tensor\sigma}_{\rm vis}.  
\ena 
Here,   $\tensor{\Pi}=  \{ \Pi_{\alpha\beta}\}$ is 
  the stress tensor in Eq.(10) determined by $n_i$,   
and $U_i$ is the wall potential. The second line 
follows from Eq.(12). The  $\tensor{\sigma}_{\rm vis}= 
\{ \sigma_{\alpha\beta}\}$ is the viscous stress  tensor of the form, 
\be 
\sigma_{\alpha\beta}= {\eta}_s (\nabla_\alpha v_\beta+ \nabla_\beta 
v_\alpha)+ (\eta_b-2{\eta}_s/3) ( \nabla\cdot{\bi v}) \delta_{\alpha\beta}, 
\en
where $\eta_s $ is the shear viscosity 
and $\eta_b$ is the bulk viscosity.   
In the previous papers on dynamical DFT \cite{Lowen,Lutsko,Goddard,Donev},   
the shear viscosity was  introduced to include  the hydrodynamic 
interaction among colloidal particles.

The total free energy  ${\cal F}_{\rm tot}={\cal F}+ {\cal K}$ 
is the sum of the Helmholtz free energy functional ${\cal F}$ 
and the fluid kinetic energy $\cal K$. The latter is  written as 
\be 
{\cal K}=  \frac{1}{2} \int d{\bi r} \rho v^2.
\en 
If the total particle numbers $N_i= \int d{\bi r}n_i$ 
are fixed and $\bi v$ vanishes on the boundaries, 
our  dynamic equations yield 
\be
\frac{d}{dt}{\cal F}_{\rm tot}= -\int d{\bi r}\bigg[ 
\Lambda^{-1}|{\bi I}|^2 + \sum_{\alpha,\beta}
\sigma_{\alpha\beta}\nabla_\alpha v_\beta \bigg]\le 0.
\en  
Thus, at long times, 
 a stationary state should be realized 
with  $\nabla\mu_1=\nabla \mu_2={\bi v}={\bi 0}$. 

In  the kinetic theory of  dilute gases, 
  $\eta_s$  tends a small constant of order $(m_1 T)^{1/2}d_1^{-2}$ 
and $\eta_b$ tends to zero in the dilute limit 
 \cite{Hir}. 
These density-dependences 
have been confirmed in molecular dynamics simulations 
 \cite{Hasse,shear,bulk}. In our case, they  
are crucial for the hydrodynamics in bubble. Thus, we used 
  simple extrapolation forms, 
\bea 
&&{\eta_s}/{ \eta_0}=  0.14+ 0.86 d_1^3 n,\\
&&{\eta_b}/{ \eta_0}= ( 0.02+ 0.98 d_1^3 n)d_1^3 n,
\ena  
where  $n=n_1+ n_2$ and 
$\eta_0$ is the viscosity in liquid.

\subsection{Numerical results after liquid decompression}
\subsubsection{Method }
We  solved the above dynamic equations 
in a spherical cell with radius $L=200d_1$, where 
$\bi v$ and $\bi I$ are parallel to ${\hat{\bi r}}= 
r^{-1}{\bi r}$, so we may set 
\be 
{\bi v} = v(r,t) {\hat{\bi r}}, \quad 
{\bi I} = I(r,t) {\hat{\bi r}}. 
\en 
 As the boundary condition, we assumed  
$ v(L,t) = I(L,t)=0$ to  fix the total particle numbers in the cell. 
 The mesh length  in  time integration was $0.2 d_1$. 

 Together with the parameters in Eq.(8), 
we assumed  $m_1=3 \times 10^{-23}$ g, $m_2= (16/9)m_1$, 
$D=2\times 10^{-4}$ cm$^2/$s, and   $\eta_0= 0.89$ cP. 
Using $\eta_0$ in Eqs.(84) and (85), we measure time in units of 
\be 
\tau= m_1/d_1\eta_0=0.112~{\rm ps}. 
\en 
Then,  we have $(d_1^2 m_1/\epsilon_{11})^{1/2} =0.041 \tau$ 
as a characteristic microscopic time 
and  $R^2/D= 4.0 \times 10^6\tau$ as a typical 
 diffusion time around a bubble with radius $R=30$ nm.   
The wall potential was assumed to be attractive as 
${ U_1} =-6.5T \exp(-{s}/{1.4})$ 
and ${U_2} =-0.5T \exp(-{s}/{1.23})$ with $s= (L-r)/d_1$, where $L=200d_1$ 
is the cell radius.
Then, the wall  prefers the solvent more than the solute and no 
surface bubble appears.


For $t<0$, we prepared a stable bubble 
with radius $R_{\rm ini}=53.11d_1$ with the initial 
densities $n_1^g=8.37 \times 10^{-5}$, $n_1^\ell=1.00$, 
$n_2^g=0.207$, and $n_2^\ell=5.68 \times 10^{-3}$ 
in units of $d_1^{-3}$.
At $t=0$, we  suddenly   decreased the 
liquid density $n_1^\ell$  in  the  region 
$r>R_{\rm ini}$ to $0.900 d_1^{-3}$, 
which gave  rise to a negative liquid pressure 
about $-2.73p_0= -419$ MPa (in a very short time). 
Then, the  bubble  expanded 
exhibiting a damped oscillation. 
At $t= 5000\tau$, the bubble radius became close to 
 the final radius  $R_\infty= 98.52d_1$, but  
the solute density in gas $n_2^g$ was $0.040d_1^{-3}$ 
and  was still noticeably smaller than the 
final value  $0.060 d_1^{-3}$. 
Note that the final equilibrium state 
can be realized   with the method in Appendix C.

Our  initial condition 
and the isothermal assumption are rather 
unrealistic, but the resultant dynamical processes  are 
dramatic, providing  fundamental 
information on  the nonlinear bubble dynamics in confined geometries.

\begin{figure}
\includegraphics[width=1\linewidth]{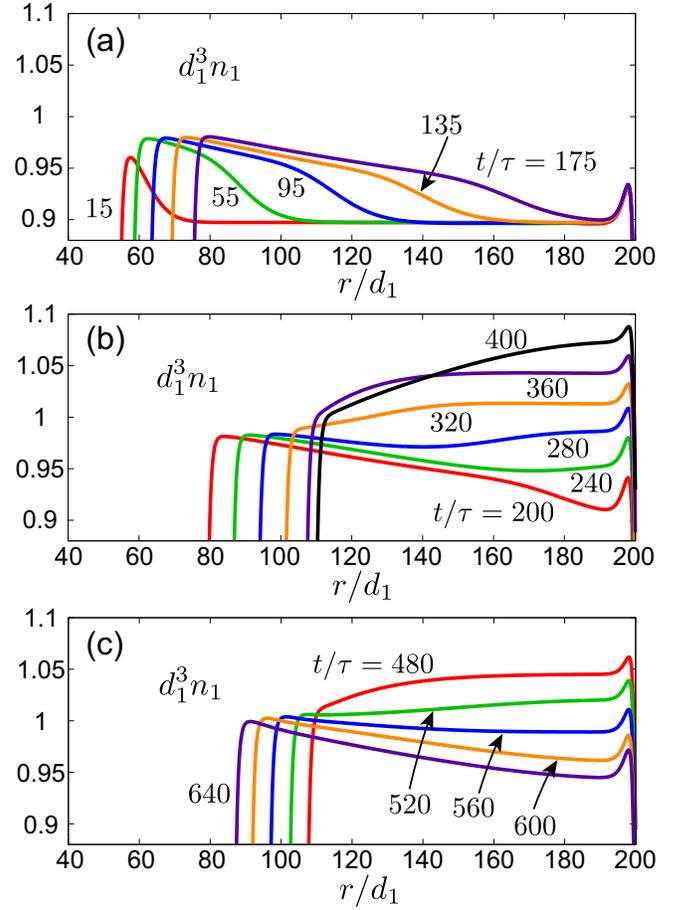}
\caption{ Time-evolution of solvent density 
$n_1(r,t)$ outside  a bubble 
with time-dependent radius $R(t)$ after decompression of liquid 
in a cell with $L=200$. 
(a) The  bubble expands emitting a large-amplitude sound  
outward. (b) Stepwise sound reaches  the cell boundary, 
while the bubble is still expanding. 
(c) The bubble is shrinking, which causes 
a large decrease of   $n_1(r,t)$ in liquid. 
 }
\end{figure}
\begin{figure}
\includegraphics[width=1\linewidth]{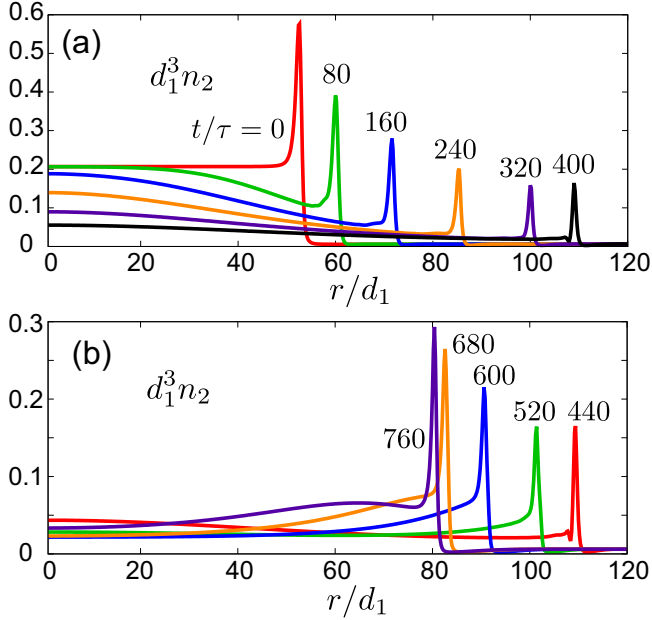}
\caption{ Time-evolution of solute density $n_2(r,t)$ inside a bubble 
during its expansion  in (a) 
and its  shrinkage in (b).  Outward and inward 
interface motions  propagate  as acoustic waves. The sound speed 
is of the same order as the interface velocity. 
In (b), scale of the vertical axis is twice  enlarged. 
 }
\end{figure}

\subsubsection{Acoustic disturbances}

In Fig.8, we show 
the profiles of the solvent density 
$n_1(r,t)$ outside the oscillating  bubble at consecutive times. 
In (a),   a large-amplitude 
acoustic wave is emitted in a stepwise form 
from the bubble surface. 
Its expanding speed is given by the (isothermal) sound velocity 
$c_\ell =\sqrt{(\p p/\p \rho)_T}= 0.88 d_1/\tau$, which is larger 
than the maximum bubble expanding speed about  $0.2 (d_1/\tau)$.    
Its amplitude is gradually decreases  away from the bubble,  
being proportional to  $r^{-1}$ in this spherically symmetric geometry. 
In (b), the wave reaches the cell boundary 
and $n_1$  increases  near the wall.  
In (c), the bubble  shrinks, 
causing an overall  density decrease in the liquid region.

In Fig.9, we also show 
the profiles  of the solute density $n_2(r,t)$  at consecutive times,
which is appreciable in the 
bubble  and has  an adsorption peak.  
The solvent density $n_1(r,t)$ remains of order $10^{-5}-
10^{-4}d_1^{-3}$ in the bubble. 
In  (a),  the outward  interface motion gives rise to 
a negative stepwise wave propagating  inward with 
the sound speed $c_g =(T/m_2)^{1/2}=0.14d_1/\tau$. Here, 
the acoustic traversal time $R/c_g $ 
is about $880\tau$ and  is close to  the 
oscillation period $650\tau$. 
In (b), $n_2(r,t)$ still 
changes inhomogeneously during the  bubble shrinkage. 
However, it becomes gradually homogeneous after 
several oscillations.  
If we would use liquid  viscosities in the bubble, 
we would have uniform dilation with $v_i \propto x_i$ 
from the early stage. It is worth noting  that 
the bubble interior has been assumed to be 
homogeneous in the literature  \cite{Plesset,Nepp,Szeri,sonol}.  
In addition, we notice that the adsorption $\Gamma$ 
in our case noticeably depends on time (see Fig.13(d) and its 
explanation).

\begin{figure}
\includegraphics[width=1\linewidth]{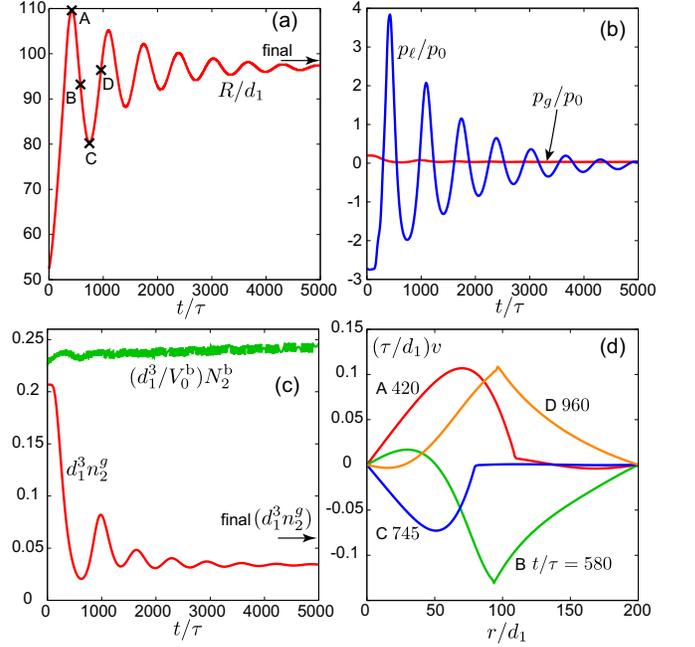}
\caption{ Earlt-stage time-evolution ($0<t<5000\tau$). 
(a) Bubble radius $R(t)$, which approaches 
the final radius (arrow). (b) Liquid  pressure 
$p_\ell(t)$ at $r=170d_1$ 
and gas pressure $p_g(t)$ at $r=2d_1$ 
 divided by $p_0= Td_1^{-3}$, 
where  $p_\ell$ exhibits a large damped oscillation 
but $p_g\cong T n_2^g$ is much smaller. 
(c) $d_1^3n_2^g(t)$ exhibiting a damped oscillation.  
 which is noticeably smaller than 
the final value(arrow). Total inter-bubble solute number $N_2^b$ 
is very slowly  increasing, so  $n_2^g(t)$ 
is nearly proportional to $ R(t)^{-3}$. 
(d) Velocity  profiles 
in the radial direction 
$v(r,t)$ divided by $d_1/\tau$ in the cell, where  
$t/\tau$ is (A) 428, (B) 580, (C) 745, and (D) 980.  
The corresponding times are marked in (a).  
}
\end{figure}
\begin{figure}
\includegraphics[width=1\linewidth]{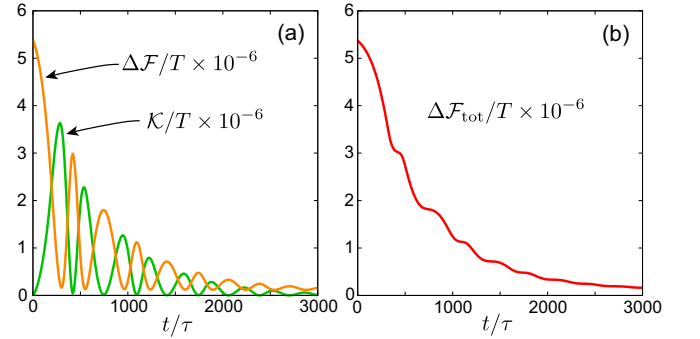}
\caption{ (a) DFT free energy deviation $\Delta{\cal F}(t)=
{\cal F}(t)- {\cal F}(\infty)$ 
and  kinetic energy ${\cal K}(t)$ vs $t/\tau$. 
(b) Their  sum  ${\cal F}_{\rm tot}(t)$.  
Here, $\Delta{\cal F}(t)$ and  ${\cal K}(t)$ 
exhibit a damped oscillation, but 
${\cal F}_{\rm tot}(t)$ decreases monotonically in time. 
  }
\end{figure}

\subsubsection{Damped oscillation}

In Fig.10, we examine early-stage 
time-evolution in the range $0<t<5000\tau=560$ ps. 
The  bubble radius 
undergoes a damped oscillation and approaches 
the final value $98.52d_1$,  
while  the particle transport through the interface 
was negligible. 
The  period is about $650\tau$ 
and the damping rate is about $5.1 \times  10^{-4} \tau^{-1}$. 
We display  (a) the bubble radius $R(t)$, which is 
defined as the peak position of $n_2(r,t)$  in our  
nonequilibrium situation (see Fig.9).  In  (b),  we plot the liquid  pressure 
$p_\ell(t)$ (taken at $r=170d_1$) 
and the gas pressure $p_g(t)$ (taken at $r=2d_1$), 
where  the former exhibits a large damped 
oscillation but the latter variation is much smaller. 
The Laplace relation does not 
hold during the bubble oscillation. 
In (c), we show  the solute density $n_2^g(t)$ 
and the total solute number within the bubble defined by 
$
N_2^b(t)=\int_{r<R} d{\bi r} n_2({\bi r},t), 
$  
Here,  $n_2^g(t)$  largely oscillates,  but 
$N_2^b(t)$  weakly increases only by a few $\%$. This 
implies    that the solute transport through the interface 
is small and  $n_2^g(t)\sim R(t)^{-3}$  in this time region. 
In addition, in  (d), we show 
 the profiles of the radial velocity  
$v(r,t)$ at four characteristic times.

In Fig.11, 
the kinetic energy ${\cal K} (t)$ in Eq.(82) and  the  
DFT free energy deviation 
$\Delta{\cal F}(t)= {\cal F}(t)- {\cal F}_\infty$ 
undergo   damped  oscillations out of phase with each other, where 
${\cal F}_\infty$ is the final value of ${\cal F}$.
However, their  sum  $\Delta{\cal F}_{\rm tot}(t)=
{\cal K}(t)+ \Delta{\cal F}(t)$  
 decreases monotonically in time in accord with Eq.(83). 
Indeed,  the radius deviation 
$\delta R$ from the mean radius (slightly smaller than $R_\infty$)
 approximately  obeys   
\be 
M{\ddot R} = -k \delta R - \zeta {\dot R}.
\en 
where ${\dot{R}}= {d \delta R}/{dt}$ and ${\ddot{R}}= {d^2\delta R}/{dt^2}$. We estimate $\zeta$ 
from $d\Delta{\cal F}/dt= -\zeta {\dot R}^2$, $k/M$ from 
the oscillation period, and $M$ from $M=2{\cal K}/{\dot R}^2$. Then, 
 $M=3.9 \rho R_\infty^3$, $k= 2.1 L/K_\ell$, and 
$\zeta= 51\eta_0 R_\infty$. Here, $\Delta{\cal F}$ 
arises from the  change in the liquid volume, while 
it is given by the surface free energy 
($\sim 4\pi \sigma R^2$)  for $L\gg R$ 
\cite{Plesset,Nepp,Szeri,sonol}. 

\begin{figure}
\includegraphics[width=1\linewidth]{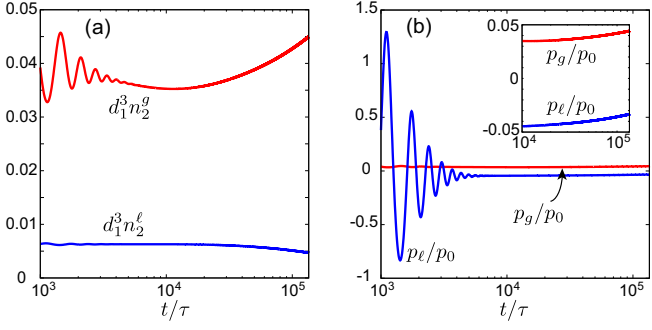}
\caption{ Long-time evolution ($t <1.3\times 10^5\tau$).
 (a)  $d_1^3n_2^g(t)$ and $d_1^3n_2^\ell(t)$   
vs $t/\tau$. For $t\gs 10^4\tau$, $n_2^g(t)$ 
 increases  and $n_2^\ell(t)$ decreases 
slowly due to a  solute flux into the bubble. 
 (b) $p_\ell(t)/p_0$ and $p_g(t)/p_0$. 
Laplace law holds after the damped oscillation. 
They increase very slowly  for $t\gs 10^4\tau$ (inset) 
due to a small increase in $R(t)$.   
 }
\end{figure}

\begin{figure}
\includegraphics[width=1\linewidth]{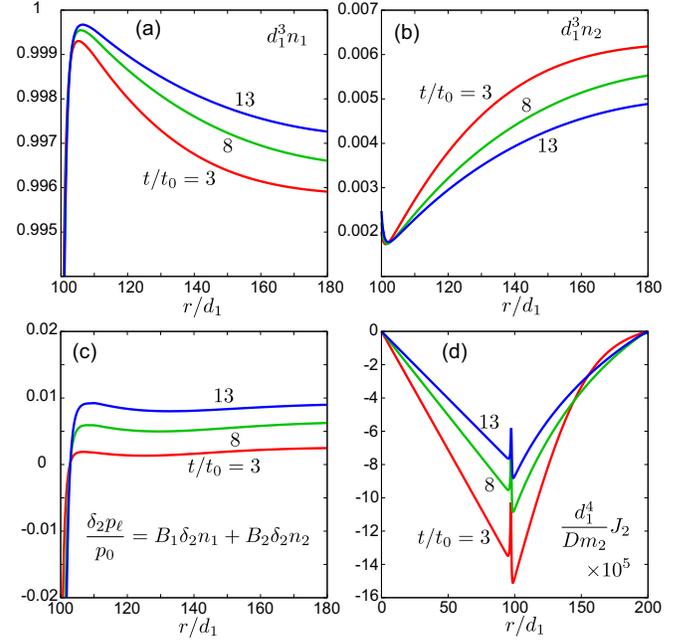}
\caption{ Profiles  at $t/t_0 = 3$, 8, and 13 
with $t_0= 10^4\tau$. (a)  $d_1^3 n_1(r,t)$ 
and (b) $d_1^3 n_2(r,t)$ in liquid. 
(c) Linear combination $\delta_2  p_\ell/p_0
\equiv B_1\delta_2 n_1+ B_2\delta_2 n_2$, where 
$\delta_2 n_i=n_i(r,t+t_0) -n_i(r,t_0)$ 
with $B_1= 40.0d_1^3$ and $B_1= 36.7 d_1^3$.  
Here, $\delta_2  p_\ell$ 
 is equal to the pressure change 
$p(r,t+t_0)- p (r, t_0)$ from Eq.(40) 
and is homogeneous away from the interface. 
(b) Solute flux $J_2= \rho_2 v+ I(r,t)$ multiplied by  $10^5 d_1^4/Dm_2$. 
 }
\end{figure}

\subsubsection{Long-time behavior}

For $t>5000\tau$, the bubble radius $R(t)$ is  close to the final one 
$R_\infty$, but small $n_2(r, t)$ still evolves diffusively 
in the liquid region. The final state should be reached 
  on a  timescale of $10^7 \tau$. 
In our case,  $K_\ell \cong 0.025/ Tn_1^\ell$ 
 and  $L\cong  2R$, so that 
   the compression pressure $\phi/K_\ell \sim 5p_0$ is much larger 
than the Laplace one $2\sigma/R  \sim 0.1 p_0$, 
where $p_0=Td_1^{-3}$. 
Moreover,  from Eq.(40),  the small pressure 
deviation  remaining in  bulk liquid is written as  
\be 
\frac{p- p_{\infty}}{ p_0}= B_1 [n_1(r,t)-n_{1\infty}] + 
 B_2 [n_2(r,t)-n_{2\infty}],
\en 
where $p_{\infty}$, $n_{1\infty}$, and  $n_{2\infty}$ 
are the final homogeneous values of $p$, $n_1$,  and $n_2$ 
in liquid, respectively, 
$B_1= 1/T K_\ell= 40.0 d_1^3$,  
and $B_2= d_1^3 [ 1+ n_1 \nu_s'(n_1)]= 36.7d_1^3$. 
Since $p$ should be homogeneous in liquid without sounds, 
this  relation indicates that
weak inhomogeneity of $n_1$ should be  induced 
by that of $n_2$ in liquid.

In Fig.12, we plot  
 $n_2^g(t)$ at $r=2d_1$ and $n_2^\ell(t)$ at $r=170d_1$ for $t <2\times 10^5\tau$.  In (a), 
we can see  a slow increase in $n_2^g(t)$  and 
a slow decrease in  $n_2^\ell(t)$ for $t\gs 10^4\tau$, 
 due to the solute transport  into the bubble. 
In  (b),   $p_\ell(t)$ and $p_g(t)$ satisfy 
the Laplace law after the damped oscillation. 
For $t\gs 10^4\tau$, they increase slowly with the fixed difference, 
 because of a small increase in  $ R$. 
Here,  $R$ increases by  $0.3d_1$ 
in time interval $[3t_0, 13t_0]$, 
producing a   compressional pressure increase about  
$0.0135p_0$. 
 

In Fig.13, we display  profiles  
at $t/t_0 = 3$, 8, and 13 
with $t_0= 10^4\tau$. In (a),   $ n_1(r,t)$ 
exhibits inhomogeneity in accord with Eq.(89). 
Its overall increase is induced by the above-mentioned 
small increase in $R$.  
In (b),   $n_2(r,t)$ relaxes 
diffusively  far from the interface. 
In (c), we examine 
the incremental changes 
$\delta_2 p_\ell\equiv  p(r,t+t_0)- p (r, t_0)$ and  
$\delta_2 n_i\equiv n_i(r,t+t_0) -n_i(r,t_0)$ 
in time interval $[t_0, t_0+t]$.  
Then, the linear combination 
$\delta_2  p_\ell/p_0
\equiv B_1\delta_2 n_1+ B_2\delta_2 n_2$ 
is surely homogeneous far from the interface.

In Fig.13(d), we display  the 
solute flux $J_2(r,t) = \rho_2 v+ I$ in the radial direction 
(see Eq.(86)). Here, $v$ is nonvanishing only in the 
 bubble interior, where $\rho_2=m_2n_2$ 
is uniform so that the gas is  dilated by   
\be 
v(r,t)= ({\dot R}/{R})r.
\en  
At the interface $r=R$ (defined as the peak position of $n_2(r,t)$), 
$J_2(r,t)$ is slightly   discontinuous across the peak of $n_2(r,t)$. 
This discontinuity gives rise to a change in  the 
surface adsorption  $\Gamma(t)$ in Eq.(57) as 
\be 
 \frac{d}{dt }\Gamma(t) = \frac{1}{m_2}[J_2(R-0,t)-J_2(R+0,t)].   
\en  
In (d), the right hand side is of order  
$10^{-5} D/d_1^4= 2.5\times 10^{-7} /d_1^2\tau$. 
In accord with this, 
$d_1^2 \Gamma$ is  0.169, 0.185, and 0.211 
at $t/t_0 =3$, $8$, and 13, respectively, 
 which indeed leads to  $d\Gamma/dt \cong 
3\times  10^{-7} / d_1^2\tau$.

\section{Summary and remarks} 

We have investigated statics and dynamics of   phase separated states   
induced  by  a neutral low-density solute using DFT.  
At fixed $T=300$ K, we have assumed   a considerably 
large solvation chemical potential $T\nu_s$  in liquid 
and a relatively weak solute-solute attractive 
interaction, under which small bubbles can appear 
in equilibrium\cite{nano,review1,review2}. 
Main results in this paper  are as follows.\\  
(i) In Sec.II, we have presented some general relations in DFT 
for the stress tensor 
$\Pi_{\alpha\beta}$ in Eqs.(12), (15), and (19) 
and for the surface tension in Eqs.(21) and (25).
The interface profiles of the densities $n_i(z)$ and the 
stress difference $\Delta\Pi(z) = \Pi_{zz}-\Pi_{xx}$ 
have   been calculated in Fig.1. These quantities have algebraic tails 
 away from the interface ($\propto |z-z_{\rm int}|^{-3}$ for Lennard-Jones potentials  as in  Fig.2). In particular, we have shown that 
the derivative $d \Delta\Pi/dz$ can be expressed in simple forms in DFT 
and in the exact statistical theory (Appendix B). 
\\
(iii) In Sec.III, we have calculated the solvation chemical potential 
$\mu_s= T\nu_s$ using  the  binary Carnahan-Starling model\cite{Car}  
in the dilute limit, as plotted in Fig.3. 
We have  then calculated solute-induced deviations, such as 
the shift of the coexisting pressure $\Delta p_{\rm cx}$,  in 
gas-liquid coexistence for dilute binary mixtures. 
We have also found  small solute-rich  bubbles, 
which are stable because they minimize 
the free energy functional of DFT. We have 
calculated the interface profiles of the densities and $\Delta\Pi$ 
for such bubbles in Fig.6. \\
(iv) In Sec.IV, we have investigated  bubble dynamics 
using dynamic equations where DFT and the hydrodynamics are 
combined. We have described a damped oscillation 
with acoustic disturbances 
 and a subsequent solute diffusion after a sudden decompression. 
In the late stage, weak  solvent inhomogeneity is 
also induced such that the liquid pressure becomes homogeneous 
as in Fig.13.\\

We make some remarks. 
(1)  We should calculate the profiles 
of surface bubbles on a wall in  various conditions. 
The dewetting transition should be  sensitive to 
a small amount of a solute (hydrophobic one for water)\cite{review2}. 
Bridging of two closely separated walls or colloidal particles by  bubbles 
is also of great importance \cite{review1,review2,Yabu}. 
(2) In our analysis, 
the solute is mildly  adsorbed at the interface due 
to a minimum of $\nu_s(n_1)$ in Eq.(38). 
 More dramatic effects of adsorption 
 should emerge with addition of  surfactants and$/$or 
 ions in the bubble formation. They are usually present in real water. 
(3)  Bubble collapse after a pressure increase 
is also worth studying. In bubble dynamics, 
we will  include  inhomogeneous 
$T$ due to adiabatic density changes 
and latent heat in the scheme of 
the dynamic van der Waals theory 
\cite{Teshi,Teshi1,Teshi2,vanderOnuki}.

\acknowledgments
This work was supported by KAKENHI No.25610122. 
One of the authors (R. O.) 
 acknowledges support from the Grant-in-Aid for Scientific 
Research on Innovative Areas ''Fluctuation and Structure'' 
from the 
Ministry of Education, Culture, Sports, Science, and 
Technology of Japan.

\vspace{6mm}
\noindent{\bf Appendix A:  Binary Carnahan-Starling model}\\
\setcounter{equation}{0}
\renewcommand{\theequation}{A\arabic{equation}}
In the binary Carnahan-Starling model\cite{Car},  the  volume fractions 
of the two species are    
\be 
\eta_i= v_{0i}n_i,\quad 
v_{0i}= \pi d_{i}^3 /6\quad (i=1,2) . 
\en 
The repulsive  part of the free energy density $ f_h$ in Eq.(1) 
is written as a function of  the total volume fraction $\eta=\eta_1+\eta_2$. 
We  here rewrite it in terms of  $u=\eta/(1-\eta)$ as 
\be
\frac{f_h}{nT}= 4u+u^2 - \frac{3}{2}
u[ 2y_1+(y_1+y_2)u] +(y_3-1)g(u). 
\en 
The  parameters $y_1$, $y_2$, and $y_3$ depend on 
$n_1$ and $n_2$ as  
\bea 
&&\hspace{-4mm} 
y_1= 
  (1+\alpha) (1-\alpha)^2n_1n_2/n(n_1+ \alpha^3 n_2), \\
&&\hspace{-4mm} 
y_2/y_1 = \alpha(n_1+\alpha^2n_2)/(1+\alpha)(n_1+\alpha^3n_2),\\
&&\hspace{-4mm}  
y_3= 
[ n_1+ \alpha^2 n_2]^3/n( n_1+ \alpha^3 n_2)^2, 
\ena 
where $n=n_1+n_2$.  In the 
one-component limit ($n_2\to 0)$\cite{Car0}, 
we have  $y_1=y_2=y_3-1=0$. 
The function  $g(u)$ in the last term in Eq.(A2) depends on $u$ as  
\be 
g(u)= u-{u^2}/{2}- \ln (1+u).
\en

The  pressure $p_h$  
from $ f_h$ in Eq.(9) is written  as   
\bea 
&&\hspace{-4mm} 
\frac{p_h}{Tn}
=(u+u^2)[4+2u -3y_1-3(y_1+y_2)u]+(1-y_3)u^3 \nonumber\\
&&\hspace{2mm} = \frac{1+\eta +\eta^2}{(1-\eta)^3}
  -\frac{3(y_1+y_2\eta)\eta  + y_3 \eta^3}{(1-\eta)^3}  -1.
\ena 
The chemical potential 
contributions $\mu_{hi}= \p  f_h/\p n_i$ from $ f_h$ 
in Eq.(6) are written as 
\bea 
&&\hspace{-6mm} \mu_{h1}=  f_h/n - X_1 +  v_{01} p_h /\eta ,\\
&&\hspace{-6mm} 
 \mu_{h2}=  f_h/n + n_1 X_1/n_2  +  v_{02} p_h /\eta ,  
\ena 
where $v_{0i}= \pi d_{ii}^3/6$.
 In terms of  the derivatives 
 $Y_i= \p y_i/\p n_1$ at fixed $n_2$, 
  $X_1$ is written as 
\be 
X_1= Tn[3u[Y_1 + (Y_{1}+ Y_{2})u]/2 - Y_{3}g(u)]. 
\en 

For small $n_2$ we have 
$y_1\cong  (1+\alpha) 
(1+\alpha)^2n_2/n$, $y_2\cong  \alpha (1+\alpha)^2 n_2/n$, and $y_3-1\cong  
(3\alpha^2-2 \alpha^3-1)n_2/n$ to linear order in $n_2$.  From Eq.(A2), $f_h$ is then expanded as 
\be 
 f_h(n_1,n_2)=Tn_1({4u_1+ u_1^2})  
+ Tn_2\nu_h (n_1)+\cdots,
\en 
where  $u_1=\eta_1/(1-\eta_1)$ 
and $\nu_h$ is given in Eq.(33). 


\vspace{6mm}
\noindent{\bf Appendix B:  Statistical-mechanical theory of surface tension and interfacial stress }\\
\setcounter{equation}{0}
\renewcommand{\theequation}{B\arabic{equation}}

We consider    binary   particle systems 
with pairwise interactions, where the  potential $\varphi_{ij}(r)$ 
includes attractive and repulsive parts. A planar interface is  placed 
perpendicularly  to the $z$ axis away from the 
walls at $z=0$ and $L$.  Then, the  average stress tensor 
can   be  expressed exactly as  \cite{Irving,Scho,Kirk,Onukibook}
\bea
&& \hspace{-5mm}
\Pi_{\alpha\beta}({\bi r}) =Tn \delta_{\alpha\beta}
- \int d{\bi r}_1 \int d{\bi r}_2 
\sum_{ij} \frac{x_{12\alpha}x_{12\beta} }{2r_{12}}\nonumber\\
&&
\times \varphi'_{ij}(r_{12}) 
\delta_s({\bi r}, {\bi r}_1,{\bi r}_2) \rho_2({\bi r}_1,{\bi r}_2),
\ena 
in terms of the $\delta_s$ function in Eq.(11). This expression  
 is analogous to that of DFT in Eq.(10). 
We introduce   the two-body  distribution function, 
\be 
\rho_{ij}({\bi r}_1, {\bi r}_2)= 
\AV{\sum_{k\neq \ell} \delta ({\bi r}_1- {\bi R}_k^i) 
 \delta ({\bi r}_2- {\bi R}_\ell^j)},   
\en  
where  $ {\bi R}_k^i$  is  the position 
of particle $k$  of species $i$.  In Eq.(B1), the  kinetic part  $Tn
\delta_{\alpha\beta}$ arises  from the Maxwell-Boltzman distribution, while  
the second term holds even in nonequilibrium.
Around a planar interface,  we may set $
 \rho_{ij}({\bi r}_1, {\bi r}_2)= \rho_{ij}(z_1, z_2,r_{12})$ 
in equilibrium.

The surface tension is  given by 
the Bakker formula in Eq.(25).  
After   integration with respect to $z$, $x_1$, and $y_1$, 
 we may remove the $\delta_s$ function to 
obtain  the Kirkwood-Buff  formula \cite{Kirk,Ono,Evansreview,Scho}, 
\be 
\sigma=   \int\hspace{-1mm}
 d{ z}_1\hspace{-1mm} \int \hspace{-1mm}
d{\bi r}_2 
\sum_{ij} \frac{r_{12}^2-3 z_{12}^2 }{4r_{12}}
 \varphi'_{ij}(r_{12}) 
 \rho_{ij}({\bi r}_1, {\bi r}_2). 
\en 
We also consider the  stress  difference 
$\Delta\Pi(z)=  \Pi_{zz}- \Pi_{xx}$ itself. 
Starting with Eq.(B1) we find    
\bea 
&&
\hspace{-5mm} 
\Delta \Pi(z)=\pi  \int_z^L \hspace{-0.5mm} dz_1\int_0^z
\hspace{-1.5mm}
 dz_2 \int_{z_{12}}^\infty 
 du \sum_{ij}(u^2/z_{12}-3z_{12})\nonumber\\ 
&&\hspace{1cm}\times{\varphi_{ij}'(u)} \rho_{ij}(z_1,z_2, u), 
\ena 
where $z_1>z>z_2$ and $z_{12}=z_1-z_2$. Use 
has been made of Eq.(18) and  the relation 
$\int dx_2 dy_2 \delta (u-r_{12})= 2\pi u \theta (u-|z_{12}|)$ at fixed $u$ 
and $z_{12}$.  
The $\sigma$  in  Eq.(B3) and the $z$ integral  of  Eq.(B4) coincide. 
 Analogously to  Eq.(28), the derivative $d\Delta\Pi/dz$ can be 
written in the double integral form,  
 \bea 
&&
\hspace{-5mm} 
\frac{d}{dz}\Delta \Pi(z)= {\pi} 
 \int_0^\infty \hspace{-1mm} d\xi  \hspace{-0.5mm} \int_{\xi}^\infty 
 du  \sum_{ij}(u^2/\xi-3\xi)\varphi_{ij}'(u) \nonumber\\ 
&&\hspace{1cm}\times [\rho_{ij}(z+\xi,z, u) -\rho_{ij}(z-\xi,z, u)]. 
\ena 
Here, if $z-z_{\rm int}\gg d_1$ 
and  $\xi-z+z_{\rm int}\gg d_1$, we can neglect the pair correlation 
at the two points and  are allowed to  replace 
$\rho_{ij}(z+\xi,z, u) -\rho_{ij}(z-\xi,z, u)$ by 
$(n_i^\ell-n_i^g)n_j^\ell \theta(\xi-z)$. Then, we obtain 
the tail in Eq.(29) if $\varphi_{ij}(r) \propto r^{-6}$ 
at long distances. 
If we replace $\varphi'_{ij}(r_{12}) 
 \rho_{ij}({\bi r}_1, {\bi r}_2)$ in Eqs.(B3)-(B5) 
by $\phi'_{ij}(r_{12}) n_i(z_1)n_j( z_2)$  and 
$Tn$ in Eq.(B1) by $Tn+p_h$, we obtain their counterparts  in DFT.

\vspace{6mm}

\noindent{\bf Appendix C:  Efficient method of calculating equilibrium states 
in DFT}\\
\setcounter{equation}{0}
\renewcommand{\theequation}{C\arabic{equation}}

As a method of minimizing  the free energy $\cal F$, 
  we seek a stationary solution of  the relaxation equations $(i=1,2)$,
\be 
\frac{\p}{\p \tau} n_i({\bi r},\tau)
= - \mu_i({\bi r},\tau) + \av{\mu_i}_{\rm cell}(\tau),
\en 
where $\mu_i $ is 
the chemical potential in Eq.(5) determined by $n_1({\bi r},\tau)$ 
 and  $\av{\mu_i}_{\rm cell}= 
\int d{\bi r}\mu_i/V$ is its  space average in the cell. 
For any initial $n_i({\bi r},0)$, 
$\mu_i ({\bi r},\tau)$ tend to be 
 homogeneous as $\tau\to \infty$. Then, 
we obtain an equilibrium state with $n_i({\bi r}) =
\lim_{\tau\to\infty}  n_i({\bi r},\tau)$. 
 We assume the one-dimensional geometry in Fig.1 
and the spherically symmetric geometry in Fig.6. 
Note that there is no physical meaning in this relaxation dynamics.


\begin{thebibliography}{99}

\bibitem{Likos}  C. N. Likos, Phys. Rep. {\bf 348}, 267 (2001).

\bibitem{Hansen} 
J. Dzubiella  and J.-P. Hansen, J. Chem. Phys. {\bf 121}, 5514 (2004). 

\bibitem{Hop} P. Hopkins, A. J. Archer, and R.  Evans, 
J. Chem. Phys. {\bf 131}, 124704 (2009). 


\bibitem{Paul} 
H.S.   Ashbaugh  and  M. E. Paulaitis,  
 J. Am. Chem. Soc. {\bf  123}, 10721 (2001). 

\bibitem{Chandler} 
D. Chandler, Nature, {\bf  437}, 640 (2005). 

\bibitem{Garde}
S. Rajamani, T.M.  Truskett, and S. Garde, 
 Proc. Natl. Acad. Sci. U.S.A. {\bf  102}, 9475 (2005).

\bibitem{Es} 
D.Beysens and D.Esteve, 
Phys. Rev. Lett. {\bf 54},   2123 (1985). 


\bibitem{Beysens} 
 D. Beysens and T. Narayanan, J. Stat. Phys. {\bf 95}, 997 (1999).

\bibitem{Dietrich} 
F. Schlesener, A. Hanke, and S. Dietrich, J. Stat. Phys. {\bf 110}, 981 (2003).

\bibitem{Oka1}
R. Okamoto and A. Onuki, Phys. Rev. E {\bf 88},  022309 (2013). 


\bibitem{Tanaka} H. Tanaka and T. Araki, Chem. Eng. Sci.{\bf 61}.  2108 (2006).

\bibitem{Araki} T. Araki and H. Tanaka, J. Phys.: Condens. Matter, {\bf 20},
 072101 (2008).


\bibitem{Furu}
A. Furukawa, A. Gambassi, S. Dietrich , and H. Tanaka, Phys.
Rev. Lett. {\bf 111}, 055701 (2013).

\bibitem{Yabu} 
S. Yabunaka, R. Okamoto, and A. Onuki, Soft Matter {\bf 11}, 5738 (2015).


\bibitem{nano} R. Okamoto and A. Onuki, 
Eur. Phys. J. E {\bf 38}, 72 (2015).


\bibitem{review1} 
P. Attard, M. P. Moody, and J.W.G. Tyrrell, 
Physica A {\bf 314},  696 (2002).  


\bibitem{review2} 
J. R. T. Seddon, D. Lohse, W. A. Ducker, and V. S. J. Craig,  
Chem. Phys. Chem.  {\bf 13}, 2179  (2012). 
\bibitem{Bu}
A.  Onuki, R.  Okamoto, and T. Araki, 
Bull. Chem. Soc. Jpn. {\bf  84},  569 (2011). 

\bibitem{Ani} 
A.F. Kostko, M.A. Anisimov, J.V. Sengers, Phys. Rev. E
70, 026118 (2004).

\bibitem{Oka-p} R. Okamoto and A. Onuki, 
Phys. Rev. E {\bf 82}, 051501 (2010).

\bibitem{Sullivan} 
D. E. Sullivan, 
  Phys. Rev. B {\bf 20}, 3991 (1979).


\bibitem{Evansreview} R. Evans, Adv. Phys. {\bf 28}, 143 (1979).

\bibitem{Tara} P. Tarazona 
and R. Evans, Mol. Phys. {\bf 48},  799 (1983). 

\bibitem{Lutsko} J. F. Lutsko, Adv. Chem. Phys. {\bf 144}, 1 (2010).

\bibitem{Car0} 
N. F. Carnahan and  K. E. Starling,  J. Chem. Phys. 
{\bf 51}, 635 (1969). 

\bibitem{Car} 
G. A. Mansoori, N. F. Carnahan, K. E. Starling, and T. W. Leland, J.
Chem. Phys. {\bf 54}, 1523 (1971).

\bibitem{Gibbs} 
J. W. Gibbs, Collected Works (Yale University Press, New Haven, CT,
1957), Vol. 1, pp. 219-331.

\bibitem{Katz} M. Blander and J.   Katz, 
 AIChE J. {\bf 21}, 833  (1975).

\bibitem{Caupin} 
M. E. M. Azouzi, C. Ramboz, J.-F. Lenain, 
and F. Caupin, Nat. Phys. {\bf 9}, 38 (2013).

\bibitem{Langer} J. S. Langer and A, J. Schwrtz, 
Phys. Rev. A {\bf 21}, 948 (1980).
\bibitem{Onukibook} A. Onuki, {\it Phase Transition Dynamics} 
(Cambridge University Press, Cambridge, 2002). 


\bibitem{Marconi} 
 U. M. B. Marconi and P. Tarazona, J. Chem. Phys. {\bf 110},
8032 (1999).

\bibitem{Archer-Evans}
A. J. Archer and R. Evans, J. Chem. Phys. 121, 4246 (2004).

\bibitem{Lowen}
 M. Rex and H. L$\rm{\ddot{o}}$wen, Eur. Phys. J. E 28, 139 (2009).

\bibitem{Goddard} 
B. D. Goddard, A. Nold, N. Savva, G. A. Pavliotis,
 and S.  Kalliadasis, Phys. Rev. Lett. {\bf 109}, 120603 (2012).

\bibitem{Donev} 
A. Donev and E. Vanden-Eijnden, J. Chem. 
Phys.{\bf 140}, 234115 (2014).

\bibitem{vanderOnuki}
A. Onuki, Phys.  Rev. E {\bf 75}, 036304 (2007).

\bibitem{vander} 
J. D. van der Waals, Verh.-K. Ned. Akad. Wet., Afd. Natuurkd.,
Eerste Reeks {\bf 1}(8), 56 (1893); English translation 
in : J. S. Rowlinson, J. Stat. Phys. {\bf 20}, 197 (1979).

\bibitem{Teshi}
R. Teshigawara and A. Onuki, Europhys. Lett. 84, 36003 (2008).
\bibitem{Teshi1} R. Teshigawara and A. Onuki, 
Phys. Rev. E 82, 021603 (2010). 
\bibitem{Teshi2} R. Teshigawara and A. Onuki, 
 Phys. Rev. E {\bf 84}, 041602 (2011).


\bibitem{Plesset} M. S. Plesset and and  A. Prosperetti,  
Ann. Rev. Fluid Mech. {\bf 9}, 145 (1977).

\bibitem{Szeri} M. M.  Fyrillas and  A. J.  Szeri, 
J. Fluid Mech. {\bf 277},  381 (1994).  

\bibitem{Nepp} 
E. A.  Neppiras,  Phys. Rep. {\bf 61}, 159 (1980).  

\bibitem{sonol} M.  P. Brenner,  S.  Hilgenfeldt, and D. Lohse, 
Rev. Mod. Phys. {\bf 74}, 426 (2002). 
\bibitem{Cahn}  J. W. Cahn,  J. Chem. Phys. {\bf 66}, 3667 (1977).


\bibitem{Irving} 
J. H. Irving and J. G. Kirkwood, J. Chem. Phys. {\bf  18}, 817 (1949). 
\bibitem{Kirk} 
J.G. Kirkwood and F. P.  Buff, J. Chem. Phys. {\bf  17}, 338 (1949). 

\bibitem{Scho}
P. Schofield, J. R. Henderson
Proc. R. Soc. Lond. A {\bf  379}, 231 (1982).


\bibitem{Sti} F. P. Buff, R. A. Lovett, and F. H. Stillinger, 
Phys. Rev. Lett. {\bf 15}, 621 (1965). 

\bibitem{Grest}
A. E. Ismail, G. S. Grest, and M. J. Stevens, J. Chem. Phys. {\bf 125}, 
014702 (2006).


\bibitem{Bakker} G. Bakker, 
in: WIEN-HARMS'
Handbuch der Experimentalphysik VI. Leipzig: Akademische Verlagsgesellschaft
1928.


\bibitem{Ono}
S. Ono and S. Kondo, {\it 
Molecular Theory of Surface Tension in Liquids} (Springer, Berlin, 1960).

\bibitem{Baker}
J. A. Barker and J. R. Henderson 
J. Chem. Phys. {\bf 76}, 6303 (1982).

\bibitem{Hauge}
J.A. Stbveng, T. Aukrust,  and E.H. Hauge, 
Physica A {\bf 143}, 40 (1987).  

\bibitem{Onuki1}
A. Onuki,  J. Chem. Phys. 
{\bf 130}, 124703 (2009). 


\bibitem{Ben} 
A. Ben-Naim  and  Y. Marcus, 
J. Chem. Phys.{\bf  81}, 2016 (1984).
 

\bibitem{Guillot} 
B.  Guillot B. and  Y.  Guissani,
J. Chem. Phys. {\bf 99}, 8075 (1993). 

\bibitem{Pratt} 
G. Hummer, S. Garde,  A. E. Garcia,   and L. R. Pratt, 
 Chem. Phys. {\bf 258}, 349 (2000).

\bibitem{Koga} M.  Ishizaki, H. Tanaka, 
and K.  Koga, Phys.  Chem.  Chem.  Phys. 
{\bf  13},  2328 (2011). 

\bibitem{Borgis} 
S. Zhao,R. Ramirez, R. Vuilleumier, and D. Borgis, 
 J. Chem. Phys. {\bf 134}, 194102 (2011). 



\bibitem{Sander} R. Sander,  
Atmos. Chem. Phys. Discuss. {\bf 14},  29615 (2014). 

\bibitem{Smith} F. L. Smith and A. H. Harvey, 
Chemical Engineering Progress, AIChE, 
{\bf 103}, 33 (2007). 
\bibitem{Onukisurf} A. Onuki, EPL  {\bf 82}, 58002 (2008). 

\bibitem{Binder} K. Binder, Physica A {\bf 319}, 99 (2003).


\bibitem{Ox} 
V. Talanquer, C. Cunningham,  and D. W. Oxtoby, 
J. Chem. Phys. {\bf 114}, 6759 (2001). 
\bibitem{Yama}
T. Yamamoto and S. Ohnishi, Phys. Chem. Chem. Phys. 
{\bf 13}, 16142 (2011).

\bibitem{Landau}  L.D. Landau and E.M. 
Lifshitz,  {\it Fluid Mechanics} (Pergamon, 1959). 


\bibitem{Hir} J. O. Hirschfelder, C. F. Curtiss, and 
R. B. Bird,    {\it Molecular Theory of Gases 
and Liquids} (Wiley, NewYork, 1954).


\bibitem{Hasse} G.A. Fernandez, J. Vrabec, and H. Hasse, 
Fluid Phase Equilibria {\bf 221} (2004) 157 (2004). 

\bibitem{shear} 
 K. Meier, A. Laesecke, and S. Kabelac, J. Chem. Phys. {\bf 121}, 3671 (2004).

\bibitem{bulk}
 K. Meier, A. Laesecke, and S. Kabelac, J. Chem. Phys. {\bf 122}, 014513 (2005). 
\end{thebibliography}
\end{document}